\documentclass[a4paper,floatfix,superscriptaddress,pra,twocolumn,longbibliography]{revtex4-2}
\usepackage[utf8]{inputenc}
\usepackage[T1]{fontenc}
\usepackage[sc,osf]{mathpazo}
\usepackage{amsmath, amsthm, amssymb, amsfonts, mathbbol, amstext}
\usepackage{mathtools}
\usepackage{graphicx}
\usepackage{dcolumn}
\usepackage{bm}
\usepackage{bbm}
\usepackage{hyperref}
\usepackage{comment}
\usepackage{color}
\usepackage{dsfont}
\usepackage[normalem]{ulem}
\usepackage{multirow}
\usepackage{diagbox}
\usepackage{bbm}
\usepackage{comment} 
\usepackage{marvosym}
\usepackage{xcolor}
\usepackage{appendix}
\usepackage{hyperref}
\usepackage{cancel}
\usepackage{subcaption}
\usepackage{ragged2e} % for the \justifying macro
\DeclareCaptionJustification{justified}{\justifying}

\newcommand{\mean}[1]{\left<#1 \right>}
\newtheorem{theorem}{Theorem}

\newtheorem{result}[theorem]{Result}

\newcommand{\ket}[1]{\left| #1 \right\rangle}
\newcommand{\bra}[1]{\left\langle #1 \right|}
\newcommand{\braket}[2]{\left\langle #1 \mid #2 \right\rangle}

\newcommand{\stkout}[1]{\ifmmode\text{\sout{\ensuremath{#1}}}\else\sout{#1}\fi}
\newcommand{\abs}[1]{\left\lvert #1 \right\rvert}

\def\Id{\mathbbm{1}}
\DeclareMathOperator{\tr}{tr}

\DeclareMathOperator{\CHSH}{CHSH}

\begin{document}
\title{Witnessing the non-objectivity of an unknown quantum dynamics}

\author{Davide Poderini}
\affiliation{International Institute of Physics, Federal University of Rio Grande do Norte, 59078-970, Natal, Brazil}
\author{Giovanni Rodari}
\affiliation{Dipartimento di Fisica - Sapienza Universit\`{a} di Roma, P.le Aldo Moro 5, I-00185 Roma, Italy}
\author{George Moreno}
\affiliation{International Institute of Physics, Federal University of Rio Grande do Norte, 59078-970, Natal, Brazil}
\affiliation{Departamento de Computação, Universidade Federal Rural de Pernambuco, 52171-900, Recife, Pernambuco, Brazil}
\author{Emanuele Polino}
\affiliation{Dipartimento di Fisica - Sapienza Universit\`{a} di Roma, P.le Aldo Moro 5, I-00185 Roma, Italy}
\author{Ranieri Nery}
\affiliation{International Institute of Physics, Federal University of Rio Grande do Norte, 59078-970, Natal, Brazil}
\author{Alessia Suprano}
\affiliation{Dipartimento di Fisica - Sapienza Universit\`{a} di Roma, P.le Aldo Moro 5, I-00185 Roma, Italy}
\author{Cristhiano Duarte}
\affiliation{International Institute of Physics, Federal University of Rio Grande do Norte, 59078-970, Natal, Brazil}
\affiliation{School of Physics and Astronomy, University of Leeds, Leeds LS2 9JT, United Kingdom}
\author{Fabio Sciarrino}
\email{fabio.sciarrino@uniroma1.it}
\affiliation{Dipartimento di Fisica - Sapienza Universit\`{a} di Roma, P.le Aldo Moro 5, I-00185 Roma, Italy}
\author{Rafael Chaves}
\email{rafael.chaves@ufrn.br}
\affiliation{International Institute of Physics, Federal University of Rio Grande do Norte, 59078-970, Natal, Brazil}
\affiliation{School of Science and Technology, Federal University of Rio Grande do Norte, Natal, Brazil}

\date{\today}
\begin{abstract}
Quantum Darwinism offers an explanation for the emergence of classical objective features -- those we are used to at macroscopic scales -- from quantum properties at the microscopic level. The interaction of a quantum system with its surroundings redundantly proliferates information to many parts of the environment, turning it accessible and objective to different observers. But given that one cannot probe the quantum system directly, only its environment, how to witness whether an unknown quantum property can be deemed objective or not? Here we propose a probabilistic framework to analyze this question and show that objectivity implies a Bell-like inequality. Among several other results, we show quantum violations of this inequality, a device-independent proof of the non-objectivity of quantum correlations that give rise to the phenomenon we name ''collective hallucination'': observers probing distinct parts of the environment can agree upon their measurement outcome of a given observable but such outcome can be totally uncorrelated from the property of the quantum system that fixed observable should be probing. We also implement an appealing photonic experiment where the temporal degree of freedom of photons is the quantum system of interest, while their polarization acts as the environment. Employing a fully black-box approach, we achieve the violation of a Bell inequality, thus certifying the non-objectivity of the underlying quantum dynamics in a fully device-independent framework. 
\end{abstract}

\maketitle

\section{Introduction}
Understanding how the quantum information encoded into a microscopic system leads to classical features, those observed at the macroscopic scales, remains a central question in the quantum foundations. In the early days of quantum theory, the comprehension of the quantum-classical boundary relied on arguably vague notions such as wave-particle duality \cite{broglie1924xxxv}, complementarity \cite{bohr1928quantum,wootters1979complementarity} or even that of a human observer \cite{wigner1963problem}. Nowadays, the tools and concepts of quantum information offer a more well-grounded framework to address those questions.

The study of decoherence \cite{schlosshauer2007decoherence,joos2013decoherence}, for instance, shows that quantum properties, such as coherence and entanglement, are degraded due to the interaction of a quantum system with its surrounding environment, a process that becomes more noticeable the larger the quantum system is \cite{aolita2008scaling}, beautifully explaining some crucial aspects of the quantum to classical transition \cite{brune1996observing,arndt1999wave,sonnentag2007measurement}. Simply put, decoherence selects the so-called pointer states \cite{zurek2003decoherence}---natural candidates for the macroscopically observed classical states obtained after a measurement---while coherent superpositions of those are suppressed. Decoherence, however, does not solve by itself the problem of how information contained in the pointer states becomes available to different measurement apparatuses, nor how this is turned into objective information, that is, independent of observers. 

That spreading of objective information is the central topic that gave rise to the idea of quantum Darwinism \cite{zurek2003decoherence,zurek2009quantum,ollivier2004objective,ollivier2005environment,blume2006quantum,korbicz2014objectivity,QD1,QD2,QD3,QD4,horodecki2015quantum,brandao2015generic,qi2021emergent,baldijao2021emergence,ccakmak2021quantum,touil2022eavesdropping}. In quantum Darwinism,  the environment---the same entity responsible for decoherence--- is also seen as a special carrier of information about the quantum system, insofar as redundantly propagating the information of the naturally selected pointer states to many external observers. Crucially, the emergence of a classical notion of objectivity is a generic feature of quantum dynamics \cite{brandao2015generic}. Irrespective of the specific modelling for the interaction with the environment, whenever the information about the pointer states is accessible to sufficiently many observers, the evolution will gradually resemble one where a specific observable is measured by all of them.

But what if other measurements, not necessarily those related to a pointer observable, are performed? In particular, if the system-environment dynamics is not known, how can one test for objectivity or, rather, the absence of it? Those are precisely the questions we address in this work.

Building on the results of \cite{brandao2015generic}, we propose a probabilistic framework to address the question of an emergent notion of objectivity. In this probabilistic setting, we associate an observer with each part of the environment (see Fig.\ \ref{fig:sys_n_envs})), and we show that the ability of each observer to encode and retrieve classical information about a quantum system translates into the emergence of an objective value for a measurement outcome. Objectivity here ought to be understood in the sense that it reflects a sort of common knowledge among the observers---\emph{a property of a quantum system is objective when it is simultaneously agreed upon by all agents}. From that, considering a particular case of two observers, we show that the Clauser-Horne-Shimony-Holt (CHSH) inequality \cite{clauser1969proposed}, a paradigmatic Bell inequality in the study of quantum non-locality \cite{brunner2014bell}, can be also be turned into a witness of non-objectivity. 

More precisely, in our probabilistic setup, the violation of a CHSH inequality implies the phenomenon we name ``collective hallucination''. This collective hallucination means that several observers can mutually agree upon their outcome for the measurement of a given observable; still, that outcome can be completely uncorrelated from the property of the quantum system it should be related to. We also prove that if objectivity is demanded for all measurements performed by the observers in the CHSH scenario, then it implies true objectivity, reflecting not only agreement between observers but also to properties of the quantum system under scrutiny. Finally, we provide a proof-of-principle experimental realization of our framework. Employing birefringent plates placed inside a Sagnac interferometer, the temporal degree of freedom of photons gets entangled with their polarization, the first being the quantum system of interest while the latter acts as its environment within the quantum Darwinism scenario.

\section{Emergence of Objectivity in Quantum Darwinism}
\label{sec:darwinism}
In the following, we review the basic notions of quantum Darwinism. We give special emphasis to the standpoint of \cite{brandao2015generic}, where the authors prove that a well-defined notion of objectivity is a generic property of any quantum dynamics. We then move forward and prove our first result, a generalization of the findings of \cite{brandao2015generic} in a general probabilistic setting, that is, not necessarily relying on (but certainly including) quantum theory.

\begin{figure}[t!]
\centering
\includegraphics[width=.35\textwidth]{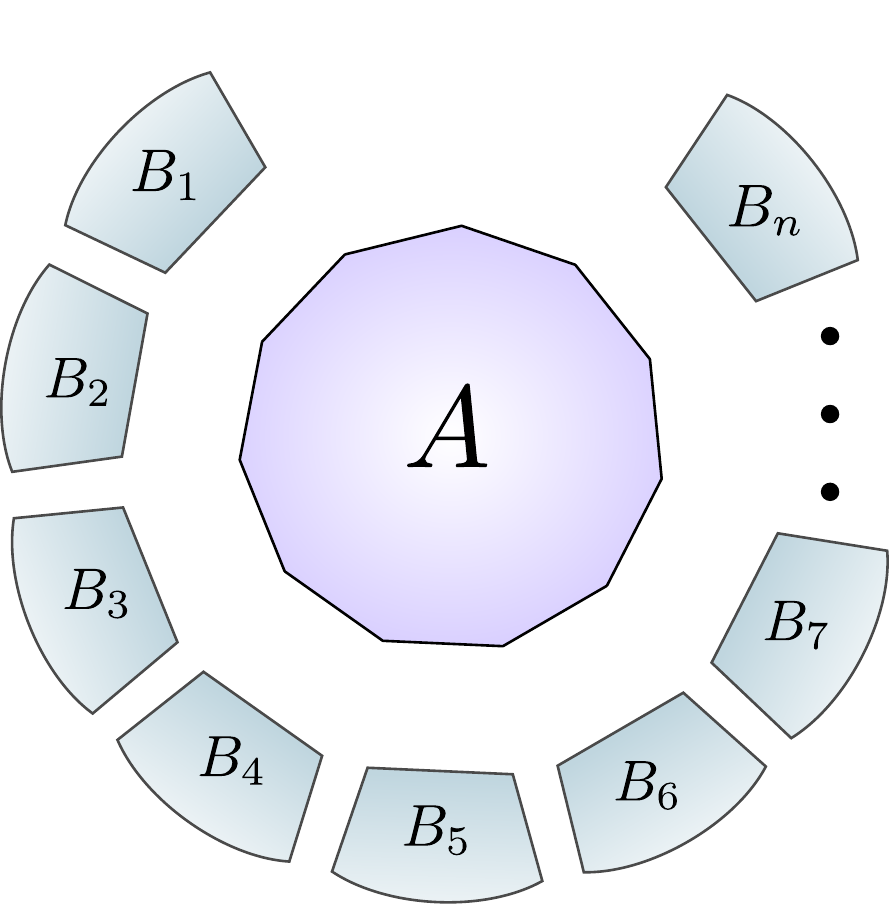}
\caption{\textbf{Quantum Darwinism scenario.} The figure depicts the general scenario considered in quantum Darwinism: one central system $A$ interacts with the environment described by $n$ systems $B_1,\ldots,B_n$. As a result of this interaction, part of the information contained in $A$ is transferred to the environment and replicated in each system $B_i$.}
\label{fig:sys_n_envs}
\end{figure}

We are interested in a general scenario where $n+1$ quantum systems interact arbitrarily, being described at a certain instant of time by a density operator $\rho_{AB_1,\dots,B_n}$---at this level of generality, it is irrelevant whether we refer to a closed system or a part of a larger system. The sub-system $A$ describes the quantum system of interest, and $B_i$ stands for the different fractions of the environment. Each fraction $B_i$ interacts with $A$ and also possibly among themselves in such a way that the quantum information originally contained in $A$ can be redundantly spread over the joint system. In a quantum description of the process, this information spreading is represented by a completely positive and trace preserving (CPTP) map $\Lambda: \mathcal{D}(A) \rightarrow \mathcal{D}(B_1 \otimes \dots \otimes B_n)$ where $\mathcal{D}(A)$ is the set of density operators on the Hilbert space associated to system $A$ (similarly to the $B_i$'s). The scenario is illustrated in Fig.~\ref{fig:sys_n_envs}

Within this context, Ref. \cite{brandao2015generic} makes a distinction between two notions of objectivity, that of observables and that of outcomes. The former states that the observers should extract information about the same observable of the system by probing parts of the environment, which would be associated with the "pointer basis" selected by the system-environment interactions. The latter considers that not only should the observable be the same, but also the value of the measurement outcome should be agreed upon by the observers.

Regarding the objectivity of observables, it follows that, quite generally, the map $\Lambda$ can be well approximated by a measure-and-prepare map, such that the reduced map for most subsets of observers is given by
\begin{equation}
\label{eq:mapm}
    \mathcal{E}_{\mathcal{B}}(\rho_A) = \sum_k \tr[\rho_A\,F_k]\,\sigma_{\mathcal{B}}^{k},
\end{equation}
where $\mathcal{B}$ is the subset of observers (or degrees of freedom of the environment being observed), $\{F_k\}_k$ is a POVM which \emph{should be the same} for all subsets $\mathcal{B}$ of the same size, $\sigma_{\mathcal{B}}^{k}$ is the (joint) quantum state for the observers in $\mathcal{B}$, prepared according to the outcome $k$ of $F_k$ and $\rho_A= \mathrm{Tr}_{B_1\dots,B_n}(\rho_{A,B_1,\dots,B_n})$. More precisely, the results in \cite{brandao2015generic} $(i)$ provide an upper bound for how close a family of measure-and-prepare maps sharing the same POVM are to the true reduced evolution $\mathcal{E}_{\mathcal{B}}$ of smaller portions $\mathcal{B}$'s and $(ii)$ show that for a suitable fraction of the observers and for a large enough number of total observers, the bound gets closer to zero, meaning that all observers would agree they are obtaining information about the same property of $\rho_A$, determined by the observable described via the POVM $\{F_k\}_{k}$.

Regarding the objectivity of outcomes, Ref. \cite{brandao2015generic} introduces the guessing probability of the outcome $k$ obtained with $F_k$ for all observers in the subset $\mathcal{B}$---tacitly assuming that the dynamics for each environment's fraction of interest has exactly the form of Eq.~\eqref{eq:mapm}.  
Consider a distribution $\{p_i\}$ and a set of states $\{\sigma_i\}$ for $i \in \{1,\ldots,m\}$ and
let $p_\mathrm{guess}(p_i, \sigma_i)$ be the guessing probability defined as:
\begin{equation}
   p_\mathrm{guess}(p_i, \sigma_i) = \max_{\{E_i\}} \sum_i p_i \tr(E_i \sigma_i),
   \label{eq:pguess}
\end{equation}
representing the capability of the ensemble of states $\{\sigma_i\}_{i \in [m]}$ to properly encode $m$ classical states distributed according to $\{p_i\}_{i \in [m]}$.
It follows that if there exists a positive $0 < \delta < 1$ such that for every observer $B_k$, with $k \in \{1,\ldots,n\}$,
    \begin{equation}
       \min_{\rho \in \mathcal{D}(A)} \left\lbrace  p_\mathrm{guess}\left(\tr(F_i \rho), \sigma_{B_k}^i\right)\right\rbrace \ge 1-\delta,
       \label{eq:bound_pguess_encoding}
    \end{equation}
then there exists some POVM $\{E^k_i\}$ for each $B_i$ such that
    \begin{equation}
       \min_{\rho \in \mathcal{D}(A)} \sum_i \tr(F_i \rho) \tr\left[ \left(\bigotimes_{k=1}^{n} E^k_i \right) \sigma^i_{B_1,\ldots,B_n} \right] \ge 1- 6n\delta^{1/4},
       \label{eq:bound_oout}
    \end{equation}
where $\{F_i\}$ is an appropriate POVM and $\sigma^i_{B_k}$ the density matrix relative to the party $B_k$ only. Qualitatively, Eq.~\eqref{eq:bound_pguess_encoding} combined with Eq.~\eqref{eq:bound_oout} show that if each $B_i$ is capable of properly encoding the outcomes of a measurement on the $A$ system, then one can assign an objective value to it, shared by all the $B_i$, in the sense that experimenters probing each single $B_i$ will all get the same value, with high probability.

Our first goal is to extend this notion of objectivity beyond quantum mechanical algebra machinery and instead rely on a purely probabilistic approach \footnote{A possible route to generalization of quantum Darwinism to GPTs was proposed in \cite{baldijao2022quantum}.
Unfortunately, the authors are concerned with defining what an idealized quantum Darwinism process would look like in GPTs, more precisely, a general version of a fan-out gate, and do not consider the noisy version of such a process.}. There are two main reasons for that. The first is that properties often seen as inherently quantum mechanical are, in fact,  also features of generalized probability theories, including monogamy of correlations \cite{barrett2006maximally} and the impossibility of broadcasting information \cite{barnum2007generalized}, just to cite a few. Understanding informational principles in such generalized settings often lead to deeper insights about quantum theory itself \cite{popescu1994quantum,MM16}. The other main reason for our approach is of practical relevance and relates to what is often called the device-independent approach to quantum information \cite{pironio2016focus}, the paradigmatic examples of which are Bell inequalities violations, non-contextuality inequality violations and their use in cryptographic protocols \cite{Ekert1991,pironio2010random,CabelloEtAl11}. In the device-independent setting, one can reach non-trivial conclusions about the quantum states being prepared or the measurements being performed by simply relying on the classical information obtained by measurement outcomes---without resorting to a detailed description of the experimental apparatus. In the particular case of quantum Darwinism, as we will see, it will allow us to not only define the concept of objectivity irrespective of any underlying dynamics or measurement setups, but also derive testable constraints on whether the statistics observed in the experiment can be deemed objective or not.

In our proposed setting, each agent $i$ has access to a portion $B_i$ of the environment surrounding $A$. Additionally, each agent $i$ is free to independently choose to measure one out of many possible observables $x_i \in \{x_{i}^{1},x_{i}^{2},...,x_{i}^{m_{i}}\}$, obtaining the corresponding outcome $b_i \in \{b_{i}^{1},b_{i}^{2},...,b_{i}^{o_{i}}\}$. If we focus only on the aggregated statistics involved in this process, the scenario is thus described by a joint probability distribution 
\begin{equation}
p(b_1,\dots,b_n\vert x_1,\dots,x_n)=\sum_{a}p(a,b_1,\dots,b_n\vert x_1,\dots,x_n),
\end{equation}
where $a$ is the outcome one would observe if a direct measurement of the system $A$, that measurement corresponding to the pointer-state observable (assuming it exists) defined by a given dynamics, had been performed. Each $x_i$ represents the random variable parametrizing the choice of which observable the $i-$th agent having access to the portion $B_i$ of the environment measures in a given run of the experiment.

According to the Born rule, a quantum description of the same scenario is given by
\begin{eqnarray}
\label{eq:jointprob}
& & p(a,b_1,\dots,b_n\vert x_1,\dots,x_n)= \\ \nonumber
& & \mathrm{Tr}[(F_a\otimes E^{1,x_1}_{b_1} \dots  \otimes E^{n,x_n}_{b_n})\,\rho_{A,B_1,\dots,B_n}],
\end{eqnarray}
where $\rho_{A,B_1,\dots,B_n}$ is the density operator representing the quantum state shared by all the environments $B_{i}$'s plus the central system $A$, and where $\{E_{b_{i}}^{i,x_{i}}\}_{b_i}$ is the POVM representing a possible choice of measurement that the $i-$th agent can realize on her fraction of the environment. It is exactly Eq.~\eqref{eq:jointprob} that motivates a general probabilistic description where the joint distribution $ p(a,b_1,\dots,b_n\vert x_1,\dots,x_n)$ has to fulfil three natural assumptions. 

The first, called \emph{no-superdeterminism} states that
\begin{equation}
\label{eq:nsdcondition}
  p(a\vert x_1,\dots,x_n)=p(a),    
\end{equation}
for every $i \in [n],$ and  for every $ x_{i} \in \{x_{i}^{1},x_{i}^{2},...,x_{i}^{m_{i}}\}$. 
In other words, the choices of which observable to measure can be made by each agent independently of how the $A$ system has been prepared or which are the pointer observables defined by a given dynamics. This is reminiscent of the measurement independence (also called ''free-will'' assumption) in Bell's theorem \cite{hall2016significance,chaves2021causal}.

The second assumption, named \emph{no-signalling}, states that 
\begin{equation}
\label{eq:nscondition}
    p(b_i \vert a,x_1,\dots,x_n)=p(b_i \vert a,x_i),
\end{equation}
for all $i \in [n]$ and for all $b_{i} \in \{b_{i}^{1},b_{i}^{2},...,b_{i}^{o_{i}}\}$. This is analogous to the no-signalling condition in Bell's theorem and implies that the choice of observables made by a given observer should not have any direct causal effect on the statistics of all other observers.

Our  final assumption, which we name \emph{$\delta$-objectivity}, is structured as follows. Let $\delta > 0$ represent an error parameter. For each agent $i$, denote $x^{*}_i$ their choice of measurement corresponding to the case where their outcome should be correlated with the outcome $a$. In a quantum description, that would precisely correspond to the pointer-state observable on system $A$, that is, corresponding to a POVM $\{E_k\}_{k}$ reproducing as reliably as possible the observable $\{F_k\}_{k}$ emerging in the effective measure-and-prepare dynamics in Eq. \eqref{eq:mapm}.
The outcome $b_i$ is $\delta$-objective, if for each observer, we have that
\begin{equation}
\sum_a p(a)\,p(b_i=a | a, x_i^*) \geq 1-\delta.
\label{eq:objcondition}
\end{equation}
The fact that this assumption brings a clearer notion of objectivity will become justified after our first result below.
For now, notice that as there is always a POVM attaining the optimal value for the guessing probability, we can create a parallel involving Eq.\eqref{eq:jointprob}, the equation defining  $p_\mathrm{guess}$, and the quantity in Eq.\eqref{eq:bound_oout} as shown below:
\begin{widetext}
\begin{align}
    p_\mathrm{guess}(\tr(F_i \rho_A), \sigma_i) = \max_{\{E_i\}} \sum_i \tr(F_i \rho_A) \tr(E_i \sigma_i)
    &\longleftrightarrow
    \sum_a p(a)p(b_i=a | a, x^*) 
    \\
    \min_{\rho_{A}} \sum_i \tr(F_i \rho_{A}) \tr\left( \bigotimes_k E^k_i \sigma_i^{1,\ldots,n} \right)
    &\longleftrightarrow
    \sum_a p(a)p(b_1=b_2=\cdots=b_n=a | a, x_1^*,\ldots,x_n^*) 
\end{align}
\end{widetext}
With that, we can state our first result, proven in Appendix~\ref{sec:app_proof1}, justifying our $\delta$-objectivity assumption.
\begin{result}
    If there exists a positive $\delta \le 1$ such that for every $k \in \{1,\ldots,n\}$:
    \begin{equation}
        \sum_a p(a)p(b_k=a | a, x_k^*) \ge 1-\delta.
    \end{equation}
    Then we have
    \begin{eqnarray}
        & \sum_a p(a)p(b_1=b_2=\cdots=b_n=a |a, x_1^*&,\ldots,x_n^*) \nonumber \\ \nonumber & & \ge 1-n\delta
    \label{eq:bound_oout_friends}
    \end{eqnarray}
\end{result}

\textbf{Remark:} Result 1 says that a result analogous to Eq. \eqref{eq:bound_oout} continues to hold, even in a fully probabilistic setting. Put another way, the inequality $\sum_a p(a)p(b_1 = b_2 = ... = b_n = a|x_{1}^{*},...,x_{n}^{*}) \geq 1 -n \delta$ expresses the possibility of assigning an objective nature to the outcome obtained by each observer. Recall that objectivity here means that regardless of the outcome obtained by each agent, that outcome is agreed upon among all the $B_i$'s, that is, $p(b_1=b_2=\cdots=b_n|x_1^*,\ldots,x_n^*)=1$. What is more, it also  reflects a property related to an observable described by a POVM $\{F_k\}_k$ acting on the subsystem $A$. In particular, when there is a perfect local agreement
(i.e., when $\delta=0$, implying $\sum_a p(b_k=a|x_k^*)=1$ for every agent), Result 1 guarantees that $\sum_a p(b_1=b_2=\cdots=b_n=a|x_1^*,\ldots,x_n^*)=1$. One can read this implication as saying that perfect local agreement implies perfect global agreement. 

\section{Bell-like inequalities witnessing non-objectivity}

The conditions of no-superdetermism, no-signalling and $\delta$-objectivity, eqs. \eqref{eq:nsdcondition}, \eqref{eq:nscondition} and \eqref{eq:objcondition} respectively, clearly define a notion for objectivity of outcomes in the probabilistic setting. Notwithstanding, note that those conditions involve the outcome $a$ that by assumption is not directly observable, as any information about it can only be obtained indirectly, by correlations of it with the outcomes $b_i$'s. Thus, similarly to Bell's theorem, $a$ plays the role of a latent or hidden variable. However, the conjunction of assumptions \eqref{eq:nsdcondition}, \eqref{eq:nscondition} and \eqref{eq:objcondition} do imply testable constraints for the observed correlations among the outcomes $b_i$. We approach those testable constraints in this section.

For doing so, we consider the particular case of only two observers ($n=2$). Each observer has two possible measurements available to them. After each measurement, a single value, out of a list of two, is flashed out. Put another way, $x_1,x_2,b_1,b_2 \in \{0,1\}$. Moreover, we specify $x_1^*$ as $x_1=0$ and $x_2^*$ as $x_2=0$---recall that each $x_i^{\ast}$ corresponds to the special case where the outcome should be correlated with the outcome $a$. We can then state our second result.

\begin{result}
Any observed correlation $p(b_1,b_2 \vert x_1,x_2)$ compatible with the conditions \eqref{eq:nsdcondition}, \eqref{eq:nscondition} and \eqref{eq:objcondition}, fulfills the inequality
\begin{multline}
\label{eq:CHSHdelta}
\mathrm{CHSH}_{\delta,\epsilon}=\mean{B^0_1B^0_2} + \mean{B^0_1B^1_2} \\- \mean{B^1_1B^0_2} + \mean{B^1_1B^1_2} \leq 2 + 4\delta-2\epsilon,
\end{multline}
with $\mean{B^0_1B^0_2}=1-2\epsilon$ where $\mean{B^{x_1}_1B^{x_2}_2}=\sum_{}(-1)^{b_1+b_2}p(b_1,b_2\vert x_1,x_2)$ is the expectation value of the observables corresponding to inputs $x_1$ and $x_2$.
\end{result}

Notice that Eq.\eqref{eq:CHSHdelta} is a relaxed version of the CHSH inequality \cite{clauser1969proposed} with one additional constraint. In Eq.~\eqref{eq:CHSHdelta} we impose that $\mean{B^0_1B^0_2}=1-2\epsilon$
to mean that both observers are in agreement (up to a discordance factor of $2\epsilon$) whenever they decide to measure the special inputs $x_1^*=0$ and $x_2^*=0$ respectively. 

Notice that our Result 1 implies that $\delta \ge \epsilon/2$ while $\delta=\epsilon/2$ corresponds to the "darwinistic" case where the disagreement $\delta$ between the observers and the latent observable follows directly from the observable disagreement $\epsilon$ between the observers themselves. Thus, any observed value $\CHSH_{\delta,\epsilon} > 2$ implies that $\delta > \epsilon/2$, witnessing non-objectivity even in the case of non-perfect agreement between the observers ($\epsilon >0$).

Considering the case where $\delta=0$ and $\epsilon=0$, our next result shows that quantum theory can violate the $\CHSH_{0,0}$ inequality while respecting $\mean{B^0_1B^0_2}=1$. That is, the observers agree among themselves but their outcomes do not reflect a property of the system $A$ to which they assume to be fully correlated with---a phenomenon we call ''collective hallucination''.
\begin{result}
\label{result3}
Quantum theory allows a violation of $\CHSH_{0,0}$ up to the value $5/2$ while respecting $\mean{B^0_1B^0_2}=1$. In particular, the maximal violation, allows to self-test a maximally two-qubit entangled state, which at the same time certifies one bit of randomness and also implies a monogamy relation. That is, even though the observers agree among themselves, the outcome of each one of them is completely uncorrelated from system A.
\end{result}

In the following, we will discuss in more depth the consequences of these results while a detailed proof is presented in the Appendix \ref{Appendix: self-test}. Notice that the violation $\CHSH_{0,0}=5/2$ is achieved considering state
\begin{equation}
    \ket{\psi}_{B_1 B_2} = \frac{1}{\sqrt{2}}(\ket{00} + \ket{11}),
    \label{eq:q_realization_state}
\end{equation}
and choosing $B_j^0 = \sigma_z$ for $j=1,2$ and
\begin{align}
    B_1^1 &= -\frac{\sigma_z}{2} - \frac{\sqrt{3}}{2} \sigma_x\\
    B_2^1 &= \frac{\sigma_z}{2} - \frac{\sqrt{3}}{2} \sigma_x \;,
    \label{eq:q_realization_measurements}
\end{align} 
As detailed in the Appendix \ref{Appendix: self-test}, the proof that $\CHSH_{0,0}=5/2$ is the maximum quantum violation relies on the idea that together with the agreement condition $\mean{B^0_1B^0_2}=1$, it allows to self-test the maximally entangled state. Recall that the possibility of performing a self-test is a sufficient condition to ensure that the quantum probability distribution achieving $\CHSH_{0,0}=5/2$ is unique, as discussed in ref. \cite{Supic2020}. Combining that uniqueness with the convex nature of the set of quantum correlations, it thus follows that $\CHSH_{0,0}=5/2$ is the maximal possible violation. Otherwise, if there was a distribution leading to a higher violation, there would be different manners to mix it with other probability distributions  (say the ones leading to maximal violation of other symmetries of this inequality) in order to obtain two different correlations reaching $\CHSH_{0,0}=5/2$, a situation that would forbid the possibility of self-testing.

Furthermore, following the arguments of Ref.~\cite{Masanes2006}, we can state that being an extreme point of the set of quantum behaviors assures that any third part event is uncorrelated with the outcomes of the observers, i.e. it holds that any realization $a$ of some third variable $A$ is such that $p(a,b_1,b_2|x_1,x_2) = p(b_1,b_2|x_1,x_2)p(a)$. Finally, because the $\CHSH_{0,0}$ inequality is invariant under the transformation $b_1' = (b_1 + 1) \bmod 2$ (the same holds for a similar transformation of $b_2$), and the behavior leading to its maximal violation is unique, we can certify a bit of randomness \cite{Dhara2013}, either $b_1$ or $b_2$. In particular, the certification of a random bit and the fact that any third party is uncorrelated implies that the probability of guessing the outcome of one of the participants is always $1/2$.

\begin{figure*}[t!]
    \centering
    \includegraphics[width=\linewidth]{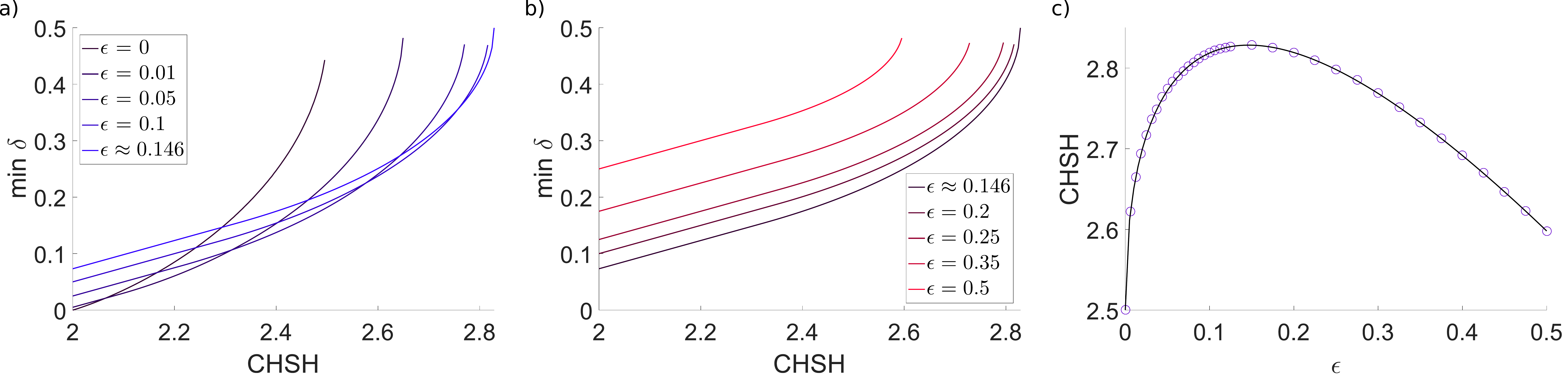}
    \caption{\textbf{a, b) Minimal values possible for $\delta$} as a function of $\CHSH_{\delta,\epsilon}$ value corresponding to the observable distribution $p(b_1,b_2|x_1,x_2)$. Results were obtained using the third level of NPA hierarchy \cite{NPA_2007}. Different curves correspond to different values for outcome agreement between observers, given by the constraint $\sum_{b_1} p(b_1=b_2|x_1=x^*,\,x_2=x^*) = 1 - \epsilon$, or equivalently $\mean{B^{0}_1B^{0}_2} = 1 - 2\epsilon$. A change in the behavior for the maximal violation of CHSH can be seen between values $\mean{B^{0}_1B^{0}_2} \leq 1/\sqrt{2}$ (shown in panel a), and values $\mean{B^{0}_1B^{0}_2} \geq 1/\sqrt{2}$ (panel b), where the former increases with increasing $\epsilon$, while the latter decreases with increasing $\epsilon$. At the same time, the restriction $\delta \geq \epsilon/2$ from Eq.\ \eqref{eq:bound_oout_friends} can be observed throughout the graphics, with saturation occurring in all cases for $\CHSH_{\delta,\epsilon}=2$. It also can be seen that a sharp rise in $\delta$ occurs near maximal violation for each $\epsilon$, which leads to numerical instabilities in these regions. For this reason, no curve reaches $\delta=0.5$ and terminal points are different for each curve. \textbf{c) Optimal values for the violation of CHSH inequality} with the constraint $\mean{B^{0}_1B^{0}_2} = 1 - 2\epsilon$, with explicit quantum realizations found numerically. The points achieve the self-testing criterion of \cite{Wang2016}, satisfying $\mathrm{asin}(\mean{B^{0}_1B^{0}_2}) + \mathrm{asin}(\mean{B^{0}_1B^{1}_2}) + \mathrm{asin}(\mean{B^{1}_1B^{0}_2}) - \mathrm{asin}(\mean{B^{1}_1B^{1}_2}) = \pi$, together with the constraints $\mean{B^{0}_1B^{1}_2} = \mean{B^{1}_1B^{0}_2} = -\mean{B^{1}_1B^{1}_2}$. The analytical curve is obtained by combining the equations, resulting in $\CHSH_{\delta,\epsilon} = 1 - 2\epsilon + 3\sin(\frac{\pi}{3} - \frac{1}{3}\mathrm{asin}(1-2\epsilon))$. 
    }
    \label{fig:Delta_CHSH}
\end{figure*}

It is worth noting that seen from Bell's theorem perspective, Result 3 also has interesting consequences for randomness certification. Differently from the usual setup that requires a violation of $\CHSH_{0,0}=2\sqrt{2}$ to certify one bit of randomness, the agreement condition $\mean{B^0_1B^0_2}=1$ permits the same to reach with a smaller CHSH inequality violation. Furthermore, the standard scenario with $\CHSH_{0,0}=2\sqrt{2}$ requires that a third input is measured by either of the parties if one wishes to directly establish a common secret bit between them. This follows from the fact that in this case, $ \mean{B^{0}_1B^{0}_2}=\mean{B^{0}_1B^{1}_2}=-\mean{B^{1}_1B^{0}_2}=\mean{B^{1}_1B^{1}_2} =1/\sqrt{2}$, that is, the measurement outcomes are not completely correlated. In turn, in our case, the condition $\mean{B^0_1B^0_2}=1$ already guarantees the perfectly correlated secret bit.

Moving beyond the case of the maximal violation $\CHSH_{0,0}=5/2$, we can also probe via a semi-definite program detailed in Appendix \ref{app:sdp}, the minimum value of $\delta$ required to explain a given value of $\CHSH_{\delta,\epsilon}$, with the results shown in Fig. \ref{fig:Delta_CHSH}. 
There we also consider the effect of imperfections on the agreement between the observers, that is, we allow $\mean{B^0_1B^0_2}=1-2\epsilon$, a condition of relevance for experimental tests of our witness. Interestingly, we observe that for any value of $\epsilon$ there is always a quantum violation of the $\CHSH_{\delta,\epsilon}$ inequality leading to $\delta =1/2$, that is, the outcomes of the observers are completely uncorrelated from the outcome $a$ of the central system they are supposedly probing. Between the range $0 \leq \epsilon \leq \frac{1}{2}(1-1/\sqrt{2})$, as we increase the $\epsilon$ we also increase the maximum quantum violation of $\CHSH_{0,0}$. From this point on, that corresponds to $\CHSH_{\delta,\epsilon}=2\sqrt{2}$ and $\mean{B^0_1B^0_2}=1/\sqrt{2}$ we see the opposite behavior, since the maximum possible violation of $\CHSH_{\delta,\epsilon}$ decreases as we increase $\epsilon$ in the range $ \frac{1}{2}(1-1/\sqrt{2}) \leq \epsilon \leq 1/2$.

Generalizing our scenario, we can now consider the case where system $A$ has not only one but actually, two properties, corresponding to the outcomes $a_0$ and $a_1$, that we assume to be correlated with the outcomes of measurements performed by the two observers. In this case, the conditions \eqref{eq:nsdcondition}, \eqref{eq:nscondition} and \eqref{eq:objcondition} now assume a joint probability distribution $p(a_0,a_1,b_1,b_2|x_1,x_2)$ and in particular, assuming $\delta=0$ for simplicity, the objectivity condition implies that
 \begin{eqnarray}
 \label{eq:objcondition2}
    \left\{\begin{array}{ll}
         p(a_i=b_1|x_1 = i, x_2)& = 1\\
         p(a_i=b_2|x_1, x_2 = i)& = 1
    \end{array}\right.
    \end{eqnarray}
Put differently, if $x_i=0$ then the outcome $b_i$ should be correlated with $a_{0}$; if $x_i=1$ then the outcome $b_i$ should be correlated with $a_{1}$. Similarly to the previous case, it follows that $\CHSH_{\delta,0}$ constraints the correlations compatible with this scenario. However, as stated in our next result, proven in the Appendix \ref{app:result4}, differently from the previous case, if we impose a stronger notion of agreement of the observers for both possible measurements, that is, $p(b_1=b_2|x_1=x_2)$, then there are no quantum violations of $\CHSH_{\delta,0}$.
\begin{result}
\label{result 4}
If we impose that $p(b_1=b_2|x_1=x_2) = 1$ for all $x_1$ and $x_2$, then all quantum correlations are compatible with the assumptions \eqref{eq:nsdcondition}, \eqref{eq:nscondition} and \eqref{eq:objcondition2}.
\end{result}

This shows that if the observers agree on the outcomes of all measurements being performed, necessarily then, the observed correlations are compatible with an underlying statistics where the observed outcomes are correlated with the property of the system they are probing, represented by the probability distribution $p(a)$.

It is worth remarking here the fact that such non-objectivity bounds can also serve as witnesses of post-quantum correlations. Differently from Result 4 for quantum correlations, no-signalling correlations, those respecting eqs. \eqref{eq:nsdcondition} and \eqref{eq:nscondition}, a set of correlations that includes the quantum set, do allow for violations even in the case where the observers agree on the outcomes of all measurements being performed. To illustrate this, it is enough to consider the paradigmatic PR-box \cite{popescu1994quantum}, given by 
\begin{equation}
    p(b_1,b_2\vert x_1,x_2 )=\frac{1}{2}\delta_{b_1\oplus b_2, x_1 \bar{x_2}} \; .
\end{equation}
The PR-box is such that the observers agree on the outcomes of both possible measurement inputs but at the same time can violate the CHSH inequality up to its algebraic maximum of $\CHSH_{0,0}=4$. In general, via our Result 4, the violation of the CHSH inequality under the constraint of concordance between the observers implies directly the post-quantum nature of the correlations.

\section{Proof-of-principle experiment: setup}
\label{sec:setup}
In the same spirit of Bell's theorem, the constraints we want to test do not depend on any specific dynamics and do not need to assume any precise physical theory. 
In the following we describe a proof-of-principle photonic experiment realizing a physical interaction dynamics whose output state can be naturally mapped into the quantum Darwinism scenario.

The quantum mechanical description of the scheme and the interaction between the involved photonic degrees of freedom provide only an intuitive picture, and we stress that the conclusions drawn from the experimental data are fully device-independent and the dynamics could in principle be unknown.

\begin{figure*}[t!]
\begin{subfigure}[t]{.47\textwidth}
\includegraphics[width=\textwidth]{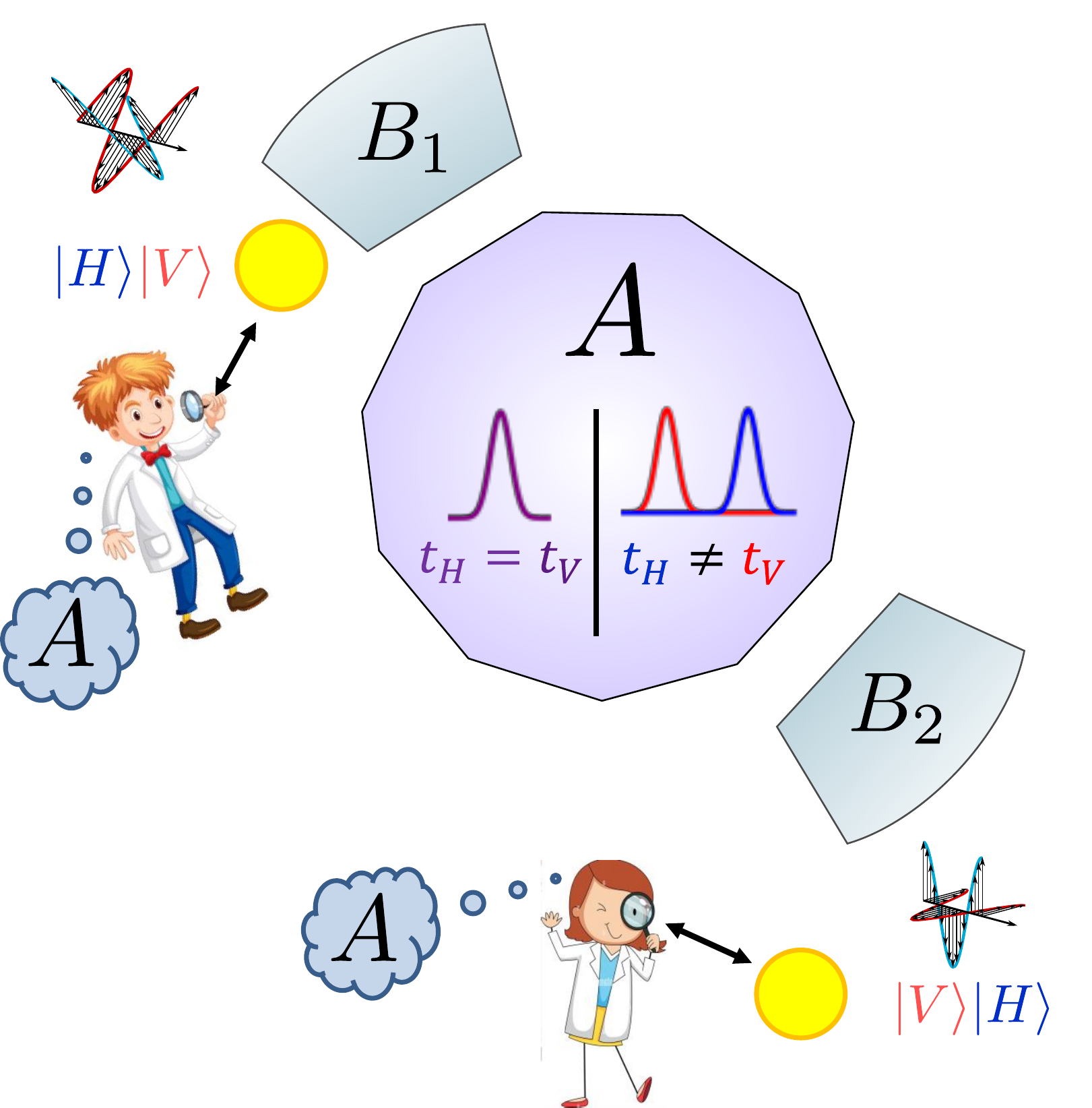}
\caption{Conceptual scheme.}
\label{fig:exp_conept}
\end{subfigure}
\begin{subfigure}[t]{.47\textwidth}
\includegraphics[width=\textwidth]{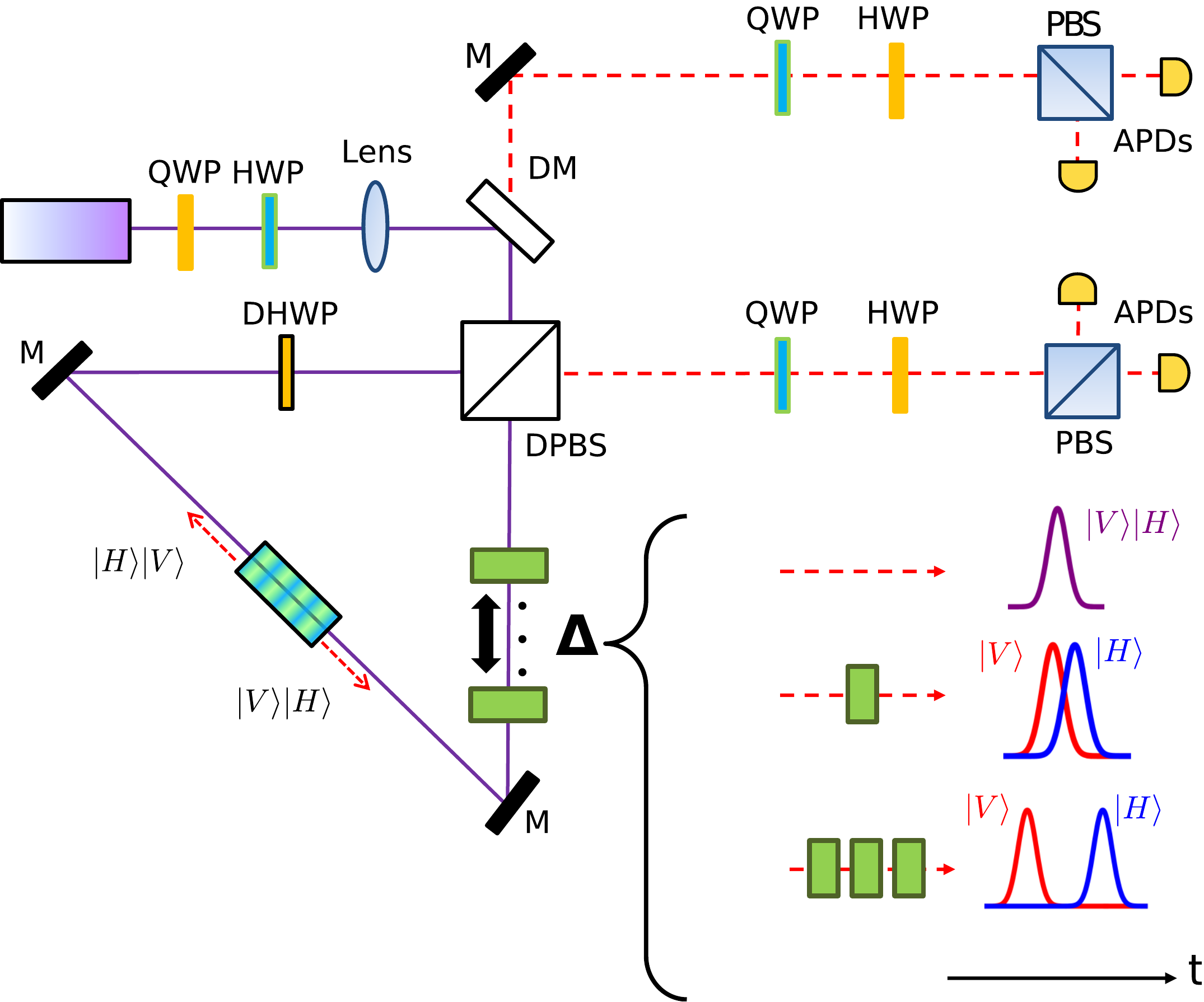}
\caption{Experimental setup}
\label{fig:exp_setup}
\end{subfigure}
\caption{\textbf{ Proof-of-principle experimental setup.}
a) Conceptual scheme of the experiment, where the polarization of photons represents the environments  while the temporal delay in the joint wavefunction represents the quantum system of interest.
b) Experimental setup.
A Sagnac-based polarization entangled photon source generates pairs of degenerate photons at $808$nm. Inside the Sagnac interferometer, the interaction of the photons generated in the anti-clockwise direction interacts with the birefringent plates which couple the polarizations with the temporal degree of freedom. The strength of the interaction is parametrized by $\Delta$ indicating the overlap between the temporal wavefunctions of the horizontal and vertical polarizations and is varied by changing the thickness of the birefringent plates. Finally, the photons are collected and detected by single photon detectors. In detail, M = mirror; PBS = polarizing beam splitter; HWP= half-waveplate; QWP = quarter-waveplate; DM = dichroic mirror; DHWP = dual-wavelength half-waveplate; APDs = avalanche photo-diode detectors.
}
\label{fig:piatte}
\end{figure*}

In our scheme, we consider the temporal degree of freedom of photons as the observed system $A$, while the polarization of two photons represents a pair of observer systems $B_1$ and $B_2$, as the environment (see  Fig.~\ref{fig:piatte}a). To simplify the presentation we consider a model where the temporal modes of the photons are treated separately, for a more in-depth analysis the reader can refer to Appendix~\ref{sec:mapping}. More specifically, two photons interact with a birefringent crystal, thus coupling the polarization and the temporal delay. Both the generation and the interaction occur within a Sagnac interferometer, after which the photons are spatially separated and their (possibly) entangled polarization carries information on the temporal delay, as shown in Fig.~\ref{fig:piatte}b.
The action $\hat{T}$ of the birefringent plate on a single photon with state $\ket{\Psi} = \alpha \ket{H}+\beta \ket{V}$ is to introduce a delay between the $\ket{H}$ and $\ket{V}$ polarization states, coupling them with a temporal wavefunction $\Phi(t)$, such that
\begin{equation}
\begin{split}
\label{eq:timeent}
\hat{T}\ket{\Psi} = \hat{T} [ (\alpha \ket{H}  +\beta \ket{V} )\, \ket{\Phi(t)}]= \\
= \alpha \ket{H} \ket{\Phi(t+\Delta t_H)} +\beta \ket{V}  \ket{\Phi(t)} \; ,
\end{split}
\end{equation}
where $\Delta t_H$ is the temporal delay, induced by the birefringent plate, on the horizontal polarization with respect to the vertical polarization.
Notice that this operation realizes a dephasing channel, a paradigmatic model to formalize decoherence in realistic physical situations, among which measurements are particular cases \cite{preskill1998lecture}. Indeed, a complex interaction involving many degrees of freedom with discrete or continuous spectrum (all realistic measurements involve several degrees of freedom of different nature) can be modeled as a unitary operation involving the considered qubit states, in our case the polarizations, and an external state corresponding to the system $A$, that in our case corresponds to the joint temporal degree of freedom. 

More specifically, the state of the qubit interacting with the birefringent plate, modeled as the dephasing channel involving time as the system $A$, can be described by the following  evolution:
\begin{equation}
\begin{split}
\label{eq:dephasing}
 &\ket{H} \otimes \ket{0}_A \xrightarrow{\hat{T}}
\Delta\ket{H} \otimes \ket{0}_A +\sqrt{1-|\Delta|^2} \ket{H} \otimes \ket{1}_A \; ,\\
&\ket{V} \otimes \ket{0}_A \xrightarrow{\hat{T}}
\ket{V} \otimes \ket{0}_A  \;,
\end{split}
\end{equation}
where $\Delta$ is the dephasing parameter related to the overlap between, respectively, the two-photon joint temporal wavefunction where the horizontal polarization is retarded with respect to the vertical polarization and the wavefunction without retardation.  The retarded states $\ket{0}_A$ and $\ket{1}_A$  are orthonormal  states of the observed system $_A\braket{i}{j}_A=\delta_{ij}$, with $i,j=0,1$ (here, $\delta$ is the Kronecker delta) corresponding to non-overlapping terms of the photonic temporal wavefunction (see Appendix~\ref{sec:mapping}).

Consider the scheme in Fig. \ref{fig:piatte}. A nonlinear crystal is placed within the interferometer and a laser pump beam enters the interferometer along an input of a dual-wavelength polarizing beam splitter (DPBS) and its interaction with the crystal generates pairs of photons in the state $\ket{HV}$. The pump passes through the interferometer along a clockwise and a counter-clockwise direction, and the relative amplitude of these contributions depends on the polarization state of the pump beam at the input of the DPBS. Note that this represents a common scheme for the generation of polarization-entangled photon pairs \cite{fedrizzi2007wavelength}. Then, a birefringent plate is inserted in a Sagnac interferometer after (seen from the anti-clockwise path) the crystal. The pump is unaffected by the plate, which induces only a negligible phase shift of the pump field, since it has a coherence time much larger than the introduced delay. On the other hand, the anti-clockwise generated $\ket{HV}$ state is affected by the dephasing channel as in \eqref{eq:dephasing}, while the clockwise term is unaffected since it does not pass through the plate. Hence, for the counter-clockwise generation, the state is given by 
\begin{equation}
\begin{split}
\label{eq:unitaryt}
 &\ket{H} \ket{V} \otimes \ket{0}_A \ket{0}_A \xrightarrow{\hat{T}}
\Delta \ket{H}\ket{V}  \otimes \ket{0}_A \ket{0}_A +  \\
&+\sqrt{1-|\Delta|^2} \ket{H} \ket{V}\otimes \ket{1}_A \ket{0}_A \;.
\end{split}
\end{equation}

Considering a pump beam in a diagonal polarization state, the (non-normalized) final state  after the DPBS of the Sagnac interferometer will be:
\begin{equation}
\begin{split}
\label{eq:unitaryt2}
 &\frac{1}{\sqrt{2}} [
\Delta \ket{H}\ket{V}  \ket{0}_A \ket{0}_A +\\
&+\sqrt{1-|\Delta|^2} \ket{H} \ket{V}\otimes \ket{1}_A \ket{0}_A-\ket{V} \ket{H}\otimes \ket{0}_A \ket{0}_A] \;.
\end{split}
\end{equation}

Tracing out the time degree of freedom, that nonetheless encodes its information in the polarization state of the photons, the final state of the two observer systems is given by
\begin{align} 
\label{eq:finalstate}
\rho_\mathrm{f}= |\Delta|^2 \ket{\Psi^-}\bra{\Psi^-}+ (1-|\Delta|^2) \rho_{\mathrm{mix}} \; ,
\end{align}
where $\ket{\Psi^-}$ is the singlet polarization state, and $\rho_{\mathrm{mix}}$ is the mixed state $\rho_{\mathrm{mix}} = (\ket{HV}\bra{HV} + \ket{VH}\bra{VH})/2$.

In this way, an interaction occurs between the polarization of the two photons and their time degree of freedom by means of the same birefringent plate. After such an interaction, the polarization of the photons, i.e. the environment systems $B_1$ and $B_2$, carry information about the time degree of freedom defined as the observed system $A$ (see Appendix~\ref{sec:mapping} for a detailed description of the mapping between the quantum Darwinism scenario and the experimental realization).  
The strength of this interaction can be tuned by changing the thickness of the birefringent plates. To illustrate this, we describe two extremal conditions. When no birefringent plate is present no interaction occurs, thus the temporal state of the photons is uncorrelated with respect to the polarization. In this case, $\Delta=1$ and, from Eq.\eqref{eq:finalstate}, the final polarization state of the photons, is a maximally entangled state $\frac{1}{\sqrt{2}}( \ket{HV}- \ket{VH})$. 

Conversely, when the thickness of the birefringent plate introduces a temporal delay greater than the coherence time of the photons, $\Delta=0$ and then the global state, from Eq. \eqref{eq:unitaryt2}, will be $(\ket{H} \ket{V}\otimes \ket{1}_A \ket{0}_A 
- \ket{V} \ket{H} \otimes \ket{0}_A \ket{0}_A)/\sqrt{2}$.
From Eq. \eqref{eq:finalstate}, tracing out the time degree of freedom, the polarization state will be the mixed state $(\ket{HV}\bra{HV}+ \ket{VH} \bra{VH})/2$.

To resume, considering the state of the photonic polarizations in Eq.\eqref{eq:finalstate}, we have that, when there is maximum coupling ($\Delta=0$) the polarization values of the two photons effectively correspond to the presence ($\ket{H}$ for the first photon and $\ket{V}$ for the second one) or absence ($\ket{V}$ for the first photon and $\ket{H}$ for the second one) of a temporal delay of the wavefunction. Conversely, when no coupling is present, no information on the presence of temporal delay is stored in the polarization of the photons. From a Quantum Darwinism perspective, polarization plays the role of an environment, mediating the interactions between the (indirectly) observed system, here the temporal delay, and the measurement apparatus.

\begin{figure*}[ht!]
\includegraphics[width=.8\textwidth]{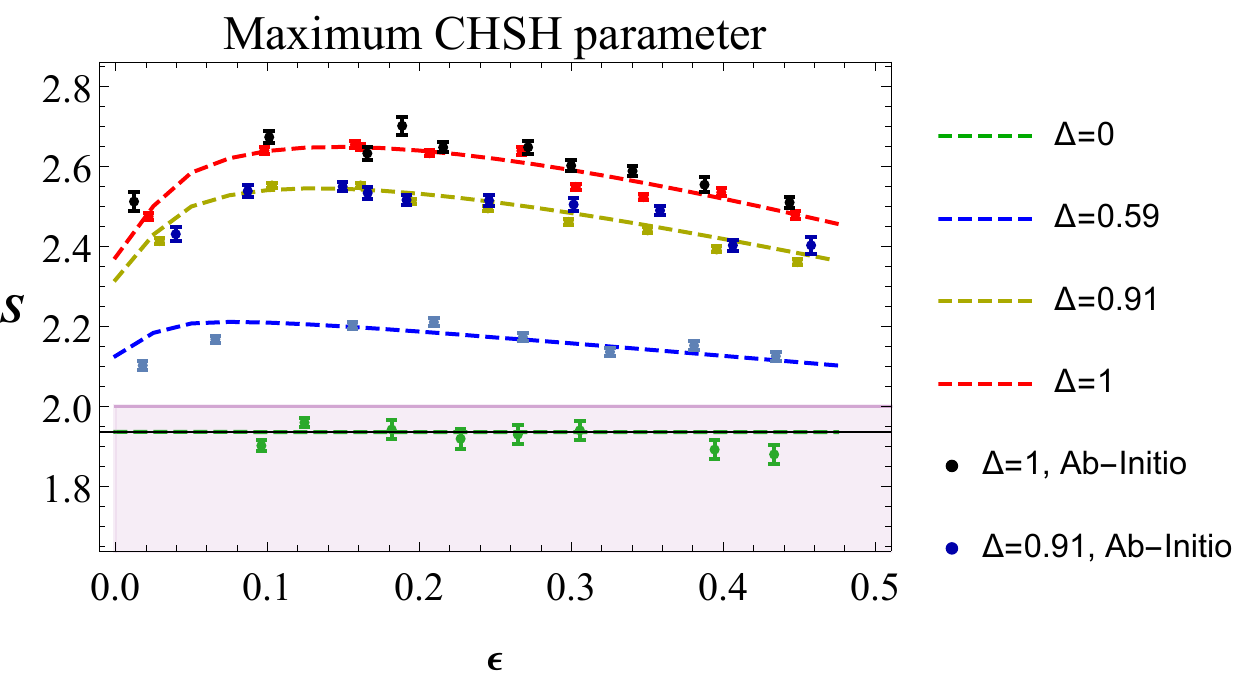}
\caption{\textbf{ Experimental data.} Optimal values of the experimental CHSH parameter as a function of the constraint $\braket{B^0_1}{B^0_2}=1-2 \epsilon$ for different values of temporal overlap $\Delta$ between the wavefunctions of the two polarizations. The dashed lines represent the optimal values calculated from the theoretical model of the experimental state. Moreover, for $\Delta=1$ and $\Delta=0.91$, there are also reported results obtained using the ab-initio approach and are indicated with the black and blue points, respectively. The error bars are calculated by assuming Poissonian statistics. 
}
\label{fig:data}
\end{figure*}

\section{Experimental results}
We performed the measurements using four different delays between the two polarizations due to the birefringent plates. For each delay, there is a correspondent strength of the interaction, parameterized by the overlap $\Delta_i$, with $i=1,2,3,4$, from Eq. \eqref{eq:unitaryt}.

Once the overlap $\Delta_i$ is fixed, the measurements are performed by varying the agreement $\mean{B^0_1B^0_2} = 1 - 2\epsilon$ between the measurements of the two observers. For each agreement, the violation of the CHSH inequality is optimized, using the information from a quantum state tomography \cite{james2005measurement}. 
More specifically, from the tomography, we extract the values of the rotation angles of the measurement waveplates able to reach, within experimental errors, the desired agreement and the corresponding maximum CHSH parameter achievable by the generated state.

The results are shown in Fig. \ref{fig:data}. 
For the case with maximum interaction, $\Delta = 0$, the polarization of the two photons became maximally entangled with the time degree of freedom and so, from the monogamy of entanglement \cite{osborne2006general}, no entanglement is possible between the polarizations. Thus, no violation of the CHSH inequality is observed  (see green points in Fig.  \ref{fig:data}). In this case, one can argue the emergence of objectivity, since the observers measuring the polarization of the two photons, not only agree among themselves but their measurement outcomes can indeed correspond to an objective property, in our proof-of-principle experiment represented by the time degree of freedom of the photons.

Conversely, when the interaction is absent ($\Delta =1$), the experimental entangled state is able to violate the CHSH inequality up to a value $S^{exp}=2.475 \pm 0.008$ with an observed agreement $\braket{B^0_1}{B^0_2}=1-2 \epsilon=0.956 \pm0.002$, that is $\epsilon=0.022 \pm 0.001$. Using the $\CHSH_{\delta. \epsilon}$ inequality \eqref{eq:CHSHdelta}, valid for general no-signalling correlations, we see that this corresponds to  $\delta \geq 0.124 \pm 0.002$. The effect becomes more pronounced when assuming the validity of quantum theory for all systems involved. A numerical computation approximating the quantum set by a superset of allowed distributions \cite{NPA_2007} returns $\delta = 0.49 \pm 0.01$, revealing that the observed variables should be almost completely uncorrelated to any candidate for the system $A$ property. We thus obtain solid experimental evidence of the (noisy) "collective hallucination", where the two observers, even if they have an agreement on the supposedly measured value, the latter cannot correspond to an objective property of the quantum system of interest.

In order to bring  our experimental evidence for non-objectivity closer to the spirit of the device-independent paradigm, we also perform measurements using the ab-initio approach introduced in \cite{poderini2022ab}, where experimental violations of classical constraints are found and optimized in a fully black-box scenario, without any knowledge the generated state and the measurement apparatuses.

More specifically, while usually in an experiment one tries to violate some inequality using precise knowledge of the employed  experimental apparatus, in an ab-initio approach one does not assume any prior information and, based only on the (noisy) output statistics, adaptively learns the optimal values of some controllable parameters, in order to optimize a given cost function, such as the violation of a Bell-like inequality.
 
In our experiment there are $8$ parameters to be optimized by the algorithm, corresponding to the values of the angles of pairs of waveplates (one pair for each measurements station) for each of the $4$ measurements needed to evaluate the CHSH parameter. In particular, the optimization firstly reaches the target value of the agreement $\mean{B^0_1B^0_2}=1-2\epsilon$ tuning the $4$ involved waveplate parameters, and then it reaches a global optimum for the CHSH value tuning the other $4$ parameters associated with the remaining CHSH measurements $\{B^1_1,B^1_2\}$. Details on the ab-initio optimization protocol can be found in Appendix \ref{sec:abinitio}. For each run of the protocol, the algorithm performs $350$ iterations, i.e. number of points sampled in a pair of $4$-dimensional parameters space associated to the CHSH measurements $\{\{B^0_1,B^0_2\},\{B^1_1,B^1_2\}\}$. The first four parameters are related to the measurements $\{B^0_1,B^0_2\}$, fixing a value for the agreement between the observers; the other parameters are related to the remaining operators $\{B^1_1,B^1_2\}$. The results on the values of the CHSH experimentally achieved with the ab-initio approach are shown in Fig.\ref{fig:data}.

The experimental points collected with the ab-initio approach can reach values higher than the ones achieved using quantum state tomography information. This is possible because within an ab-initio framework errors in the characterization of the optical setup, such as the optical axes of waveplates, can be compensated automatically by the optimization process. For all the curves with $\Delta \ne 0$, for each value of the observed agreement, the CHSH is violated, consequently witnessing a degree of non-objectivity in a device-independent way.

\section{Discussion}
Comprehending how microscopic quantum features give rise to the observed macroscopic properties is a central goal of decoherence theory and in particular of quantum Darwinism. Importantly, the emergence of objectivity, that is, the fact that different observers agree on the properties of a quantum system under observation, can be seen as a generic aspect as long as the information of the quantum system is successfully outspread to the environment it is interacting with. It is unclear, however, how to witness the presence or rather the absence of such objectivity in practice. Can we witness non-objectivity by simply probing the environment, without any knowledge of the underlying dynamics?

To answer in positive to this question, we establish a probabilistic framework to cast objectivity through operational lens, building up on the results of \cite{brandao2015generic}. Within this setting, we propose three properties defining what is to be expected from a generic objective behaviour: no-superdeterminism, no-signalling and the $\delta$-objectivity, the latter stating that
$p(b_i=a | x_i^*) \geq 1-\delta$, where $x_i^*$ denotes the measurement for which the observer should try to recover as best as possible the information about the system $A$ as encoded in the probability $p(a)$. Those conditions play a role similar to what the concept of local realism implies for Bell's theorem \cite{brunner2014bell}. In particular, the notion of $\delta$-objectivity is justified by our first result stating that the local agreement between a given observer and the quantum system of interest, translates into a global notion of agreement between all observers having access to some part of the environment.

We then showed that a generalization of the seminal CHSH inequality \cite{clauser1969proposed} constraints the set of possible correlations compatible with the three aforementioned assumptions. Furthermore, the violation of this inequality offers a device-independent witness of the non-objectivity of the underlying process, at the same time that it naturally quantifies how much one should give up objectivity in order to explain the observed correlations. Further, we proved that quantum mechanics allow for violations of this inequality and, in particular, leads to a monogamy relation implying that even though the observers agree among themselves, their outcomes are completely uncorrelated from the system they supposedly should be correlated with. A phenomenon we name "collective hallucination" and that we have experimentally probed using a photonic setup where the quantum property of interest is encoded in the temporal degree of freedom of photons, the polarization of which plays the role of the environment to which the information should redundantly proliferate. 

For scenarios where the probed system has more than one property of interest, we demonstrated that if the observers have to agree on measurement outcomes of all performed measurements, then quantum correlations are compatible with the assumptions no-superdeterminism, no-signalling and objectivity, that is, they cannot violate any Bell-like inequality. A result that can be violated by correlations beyond those allowed by quantum theory and that thus can be employed as a test for post-quantumness.

To our knowledge, this is the first connection between quantum Darwinism, and the notion of objectivity it entails, with Bell inequalities and  device-independent quantum information processing, a bridge that deserves further investigation. For instance, it would be interesting to generalize our results to a larger number of observers and consider measurements with more outcomes.
At the same time, one should understand paradigmatic dynamics considered in the literature of quantum Darwinism~\cite{ollivier2004objective, ollivier2005environment, blume2006quantum, korbicz2014objectivity, QD1, QD2, QD3, QD4} under this new perspective, and explore the connections with other objectivity measures~\cite{girolami2022redundantly} valid in the quantum framework. 
It is worth remarking also that our approach could lead to substantial refinements in recent tests for the emergence of objectivity \cite{ciampini2018experimental,chen2019emergence,unden2019}. That said, we should also note that another bridge connecting quantum Darwinism, Spekkens contextuality and quantum information has also been recently erected. Adapting the prepare-and-measure scenario into the usual environment as a witness framework, in Ref.~\cite{baldijao2021emergence}, the authors managed to prove that Spekkens non-contextuality for each observer follows through whenever the environment proliferates the information about the central system appropriately. Their notion of classicality differs from ours, insofar as here we consider mutual agreement rather than non-contextuality as a signature of classicality---and our connection with foundations of quantum mechanics is via Bell scenarios. Additionally, our work goes a step further, as we investigate our theoretical findings with a proof-of-principle experiment.    

Finally, we notice that the $\delta$-objectivity constraint we consider here is mathematically very similar to the notion of absoluteness of events employed to analyze a generalization of the Wigner's friend experiment \cite{bong2020strong,Moreno2021,wiseman2022thoughtful} in the foundations of quantum theory. Apprehending further the connections between quantum Darwinism/objectivity and Wigner's friend experiment/absoluteness of events is another relevant research direction that we hope our results might trigger.

\section{Acknowledgements} 
This work was supported by the Serrapilheira Institute (Grant No. Serra-1708-15763), by the Simons Foundation (Grant Number 884966, AF), the Brazilian National Council for Scientific and Technological Development (CNPq) via the National Institute for Science and Technology on Quantum Information (INCT-IQ) and Grants No. 406574/2018-9 and 307295/2020-6 This research was also supported by the Fetzer Franklin Fund of the John E.\ Fetzer Memorial Trust and by grant number FQXi-RFP-IPW-1905 from the Foundational Questions Institute and Fetzer Franklin Fund, a donor advised fund of Silicon Valley Community Foundation.  
We acknowledge support from the Templeton Foundation, The Quantum Information Structure of Spacetime (QISS) Project (qiss.fr) (the opinions expressed in this publication are those of the author(s) and do not necessarily reflect the views of the John Templeton Foundation)  Grant Agreement No.  61466.
\bibliography{main}

\begin{thebibliography}{60}%
\makeatletter
\providecommand \@ifxundefined [1]{%
 \@ifx{#1\undefined}
}%
\providecommand \@ifnum [1]{%
 \ifnum #1\expandafter \@firstoftwo
 \else \expandafter \@secondoftwo
 \fi
}%
\providecommand \@ifx [1]{%
 \ifx #1\expandafter \@firstoftwo
 \else \expandafter \@secondoftwo
 \fi
}%
\providecommand \natexlab [1]{#1}%
\providecommand \enquote  [1]{``#1''}%
\providecommand \bibnamefont  [1]{#1}%
\providecommand \bibfnamefont [1]{#1}%
\providecommand \citenamefont [1]{#1}%
\providecommand \href@noop [0]{\@secondoftwo}%
\providecommand \href [0]{\begingroup \@sanitize@url \@href}%
\providecommand \@href[1]{\@@startlink{#1}\@@href}%
\providecommand \@@href[1]{\endgroup#1\@@endlink}%
\providecommand \@sanitize@url [0]{\catcode `\\12\catcode `\$12\catcode
  `\&12\catcode `\#12\catcode `\^12\catcode `\_12\catcode `\%12\relax}%
\providecommand \@@startlink[1]{}%
\providecommand \@@endlink[0]{}%
\providecommand \url  [0]{\begingroup\@sanitize@url \@url }%
\providecommand \@url [1]{\endgroup\@href {#1}{\urlprefix }}%
\providecommand \urlprefix  [0]{URL }%
\providecommand \Eprint [0]{\href }%
\providecommand \doibase [0]{https://doi.org/}%
\providecommand \selectlanguage [0]{\@gobble}%
\providecommand \bibinfo  [0]{\@secondoftwo}%
\providecommand \bibfield  [0]{\@secondoftwo}%
\providecommand \translation [1]{[#1]}%
\providecommand \BibitemOpen [0]{}%
\providecommand \bibitemStop [0]{}%
\providecommand \bibitemNoStop [0]{.\EOS\space}%
\providecommand \EOS [0]{\spacefactor3000\relax}%
\providecommand \BibitemShut  [1]{\csname bibitem#1\endcsname}%
\let\auto@bib@innerbib\@empty
%</preamble>
\bibitem [{\citenamefont {Broglie}(1924)}]{broglie1924xxxv}%
  \BibitemOpen
  \bibfield  {author} {\bibinfo {author} {\bibfnamefont {L.~d.}\ \bibnamefont
  {Broglie}},\ }\bibfield  {title} {\bibinfo {title} {A tentative theory of
  light quanta},\ }\href@noop {} {\bibfield  {journal} {\bibinfo  {journal}
  {The London, Edinburgh, and Dublin Philosophical Magazine and Journal of
  Science}\ }\textbf {\bibinfo {volume} {47}},\ \bibinfo {pages} {446}
  (\bibinfo {year} {1924})}\BibitemShut {NoStop}%
\bibitem [{\citenamefont {Bohr}\ \emph {et~al.}(1928)\citenamefont {Bohr} \emph
  {et~al.}}]{bohr1928quantum}%
  \BibitemOpen
  \bibfield  {author} {\bibinfo {author} {\bibfnamefont {N.}~\bibnamefont
  {Bohr}} \emph {et~al.},\ }\href@noop {} {\emph {\bibinfo {title} {The quantum
  postulate and the recent development of atomic theory}}},\ Vol.~\bibinfo
  {volume} {3}\ (\bibinfo  {publisher} {Printed in Great Britain by R. \& R.
  Clarke, Limited},\ \bibinfo {year} {1928})\BibitemShut {NoStop}%
\bibitem [{\citenamefont {Wootters}\ and\ \citenamefont
  {Zurek}(1979)}]{wootters1979complementarity}%
  \BibitemOpen
  \bibfield  {author} {\bibinfo {author} {\bibfnamefont {W.~K.}\ \bibnamefont
  {Wootters}}\ and\ \bibinfo {author} {\bibfnamefont {W.~H.}\ \bibnamefont
  {Zurek}},\ }\bibfield  {title} {\bibinfo {title} {Complementarity in the
  double-slit experiment: Quantum nonseparability and a quantitative statement
  of bohr's principle},\ }\href@noop {} {\bibfield  {journal} {\bibinfo
  {journal} {Physical Review D}\ }\textbf {\bibinfo {volume} {19}},\ \bibinfo
  {pages} {473} (\bibinfo {year} {1979})}\BibitemShut {NoStop}%
\bibitem [{\citenamefont {Wigner}(1963)}]{wigner1963problem}%
  \BibitemOpen
  \bibfield  {author} {\bibinfo {author} {\bibfnamefont {E.~P.}\ \bibnamefont
  {Wigner}},\ }\bibfield  {title} {\bibinfo {title} {The problem of
  measurement},\ }\href@noop {} {\bibfield  {journal} {\bibinfo  {journal}
  {American Journal of Physics}\ }\textbf {\bibinfo {volume} {31}},\ \bibinfo
  {pages} {6} (\bibinfo {year} {1963})}\BibitemShut {NoStop}%
\bibitem [{\citenamefont {Schlosshauer}(2007)}]{schlosshauer2007decoherence}%
  \BibitemOpen
  \bibfield  {author} {\bibinfo {author} {\bibfnamefont {M.~A.}\ \bibnamefont
  {Schlosshauer}},\ }\href@noop {} {\emph {\bibinfo {title} {Decoherence: and
  the quantum-to-classical transition}}}\ (\bibinfo  {publisher} {Springer
  Science \& Business Media},\ \bibinfo {year} {2007})\BibitemShut {NoStop}%
\bibitem [{\citenamefont {Joos}\ \emph {et~al.}(2013)\citenamefont {Joos},
  \citenamefont {Zeh}, \citenamefont {Kiefer}, \citenamefont {Giulini},
  \citenamefont {Kupsch},\ and\ \citenamefont
  {Stamatescu}}]{joos2013decoherence}%
  \BibitemOpen
  \bibfield  {author} {\bibinfo {author} {\bibfnamefont {E.}~\bibnamefont
  {Joos}}, \bibinfo {author} {\bibfnamefont {H.~D.}\ \bibnamefont {Zeh}},
  \bibinfo {author} {\bibfnamefont {C.}~\bibnamefont {Kiefer}}, \bibinfo
  {author} {\bibfnamefont {D.~J.}\ \bibnamefont {Giulini}}, \bibinfo {author}
  {\bibfnamefont {J.}~\bibnamefont {Kupsch}},\ and\ \bibinfo {author}
  {\bibfnamefont {I.-O.}\ \bibnamefont {Stamatescu}},\ }\href@noop {} {\emph
  {\bibinfo {title} {Decoherence and the appearance of a classical world in
  quantum theory}}}\ (\bibinfo  {publisher} {Springer Science \& Business
  Media},\ \bibinfo {year} {2013})\BibitemShut {NoStop}%
\bibitem [{\citenamefont {Aolita}\ \emph {et~al.}(2008)\citenamefont {Aolita},
  \citenamefont {Chaves}, \citenamefont {Cavalcanti}, \citenamefont
  {Ac{\'\i}n},\ and\ \citenamefont {Davidovich}}]{aolita2008scaling}%
  \BibitemOpen
  \bibfield  {author} {\bibinfo {author} {\bibfnamefont {L.}~\bibnamefont
  {Aolita}}, \bibinfo {author} {\bibfnamefont {R.}~\bibnamefont {Chaves}},
  \bibinfo {author} {\bibfnamefont {D.}~\bibnamefont {Cavalcanti}}, \bibinfo
  {author} {\bibfnamefont {A.}~\bibnamefont {Ac{\'\i}n}},\ and\ \bibinfo
  {author} {\bibfnamefont {L.}~\bibnamefont {Davidovich}},\ }\bibfield  {title}
  {\bibinfo {title} {Scaling laws for the decay of multiqubit entanglement},\
  }\href@noop {} {\bibfield  {journal} {\bibinfo  {journal} {Physical Review
  Letters}\ }\textbf {\bibinfo {volume} {100}},\ \bibinfo {pages} {080501}
  (\bibinfo {year} {2008})}\BibitemShut {NoStop}%
\bibitem [{\citenamefont {Brune}\ \emph {et~al.}(1996)\citenamefont {Brune},
  \citenamefont {Hagley}, \citenamefont {Dreyer}, \citenamefont {Maitre},
  \citenamefont {Maali}, \citenamefont {Wunderlich}, \citenamefont {Raimond},\
  and\ \citenamefont {Haroche}}]{brune1996observing}%
  \BibitemOpen
  \bibfield  {author} {\bibinfo {author} {\bibfnamefont {M.}~\bibnamefont
  {Brune}}, \bibinfo {author} {\bibfnamefont {E.}~\bibnamefont {Hagley}},
  \bibinfo {author} {\bibfnamefont {J.}~\bibnamefont {Dreyer}}, \bibinfo
  {author} {\bibfnamefont {X.}~\bibnamefont {Maitre}}, \bibinfo {author}
  {\bibfnamefont {A.}~\bibnamefont {Maali}}, \bibinfo {author} {\bibfnamefont
  {C.}~\bibnamefont {Wunderlich}}, \bibinfo {author} {\bibfnamefont {J.~M.}\
  \bibnamefont {Raimond}},\ and\ \bibinfo {author} {\bibfnamefont
  {S.}~\bibnamefont {Haroche}},\ }\bibfield  {title} {\bibinfo {title}
  {Observing the progressive decoherence of the “meter” in a quantum
  measurement},\ }\href@noop {} {\bibfield  {journal} {\bibinfo  {journal}
  {Physical Review Letters}\ }\textbf {\bibinfo {volume} {77}},\ \bibinfo
  {pages} {4887} (\bibinfo {year} {1996})}\BibitemShut {NoStop}%
\bibitem [{\citenamefont {Arndt}\ \emph {et~al.}(1999)\citenamefont {Arndt},
  \citenamefont {Nairz}, \citenamefont {Vos-Andreae}, \citenamefont {Keller},
  \citenamefont {Van~der Zouw},\ and\ \citenamefont
  {Zeilinger}}]{arndt1999wave}%
  \BibitemOpen
  \bibfield  {author} {\bibinfo {author} {\bibfnamefont {M.}~\bibnamefont
  {Arndt}}, \bibinfo {author} {\bibfnamefont {O.}~\bibnamefont {Nairz}},
  \bibinfo {author} {\bibfnamefont {J.}~\bibnamefont {Vos-Andreae}}, \bibinfo
  {author} {\bibfnamefont {C.}~\bibnamefont {Keller}}, \bibinfo {author}
  {\bibfnamefont {G.}~\bibnamefont {Van~der Zouw}},\ and\ \bibinfo {author}
  {\bibfnamefont {A.}~\bibnamefont {Zeilinger}},\ }\bibfield  {title} {\bibinfo
  {title} {Wave--particle duality of c60 molecules},\ }\href@noop {} {\bibfield
   {journal} {\bibinfo  {journal} {nature}\ }\textbf {\bibinfo {volume}
  {401}},\ \bibinfo {pages} {680} (\bibinfo {year} {1999})}\BibitemShut
  {NoStop}%
\bibitem [{\citenamefont {Sonnentag}\ and\ \citenamefont
  {Hasselbach}(2007)}]{sonnentag2007measurement}%
  \BibitemOpen
  \bibfield  {author} {\bibinfo {author} {\bibfnamefont {P.}~\bibnamefont
  {Sonnentag}}\ and\ \bibinfo {author} {\bibfnamefont {F.}~\bibnamefont
  {Hasselbach}},\ }\bibfield  {title} {\bibinfo {title} {Measurement of
  decoherence of electron waves and visualization of the quantum-classical
  transition},\ }\href@noop {} {\bibfield  {journal} {\bibinfo  {journal}
  {Physical review letters}\ }\textbf {\bibinfo {volume} {98}},\ \bibinfo
  {pages} {200402} (\bibinfo {year} {2007})}\BibitemShut {NoStop}%
\bibitem [{\citenamefont {Zurek}(2003)}]{zurek2003decoherence}%
  \BibitemOpen
  \bibfield  {author} {\bibinfo {author} {\bibfnamefont {W.~H.}\ \bibnamefont
  {Zurek}},\ }\bibfield  {title} {\bibinfo {title} {Decoherence, einselection,
  and the quantum origins of the classical},\ }\href@noop {} {\bibfield
  {journal} {\bibinfo  {journal} {Reviews of modern physics}\ }\textbf
  {\bibinfo {volume} {75}},\ \bibinfo {pages} {715} (\bibinfo {year}
  {2003})}\BibitemShut {NoStop}%
\bibitem [{\citenamefont {Zurek}(2009)}]{zurek2009quantum}%
  \BibitemOpen
  \bibfield  {author} {\bibinfo {author} {\bibfnamefont {W.~H.}\ \bibnamefont
  {Zurek}},\ }\bibfield  {title} {\bibinfo {title} {Quantum darwinism},\
  }\href@noop {} {\bibfield  {journal} {\bibinfo  {journal} {Nature physics}\
  }\textbf {\bibinfo {volume} {5}},\ \bibinfo {pages} {181} (\bibinfo {year}
  {2009})}\BibitemShut {NoStop}%
\bibitem [{\citenamefont {Ollivier}\ \emph {et~al.}(2004)\citenamefont
  {Ollivier}, \citenamefont {Poulin},\ and\ \citenamefont
  {Zurek}}]{ollivier2004objective}%
  \BibitemOpen
  \bibfield  {author} {\bibinfo {author} {\bibfnamefont {H.}~\bibnamefont
  {Ollivier}}, \bibinfo {author} {\bibfnamefont {D.}~\bibnamefont {Poulin}},\
  and\ \bibinfo {author} {\bibfnamefont {W.~H.}\ \bibnamefont {Zurek}},\
  }\bibfield  {title} {\bibinfo {title} {Objective properties from subjective
  quantum states: Environment as a witness},\ }\href@noop {} {\bibfield
  {journal} {\bibinfo  {journal} {Physical review letters}\ }\textbf {\bibinfo
  {volume} {93}},\ \bibinfo {pages} {220401} (\bibinfo {year}
  {2004})}\BibitemShut {NoStop}%
\bibitem [{\citenamefont {Ollivier}\ \emph {et~al.}(2005)\citenamefont
  {Ollivier}, \citenamefont {Poulin},\ and\ \citenamefont
  {Zurek}}]{ollivier2005environment}%
  \BibitemOpen
  \bibfield  {author} {\bibinfo {author} {\bibfnamefont {H.}~\bibnamefont
  {Ollivier}}, \bibinfo {author} {\bibfnamefont {D.}~\bibnamefont {Poulin}},\
  and\ \bibinfo {author} {\bibfnamefont {W.~H.}\ \bibnamefont {Zurek}},\
  }\bibfield  {title} {\bibinfo {title} {Environment as a witness: Selective
  proliferation of information and emergence of objectivity in a quantum
  universe},\ }\href@noop {} {\bibfield  {journal} {\bibinfo  {journal}
  {Physical review A}\ }\textbf {\bibinfo {volume} {72}},\ \bibinfo {pages}
  {042113} (\bibinfo {year} {2005})}\BibitemShut {NoStop}%
\bibitem [{\citenamefont {Blume-Kohout}\ and\ \citenamefont
  {Zurek}(2006)}]{blume2006quantum}%
  \BibitemOpen
  \bibfield  {author} {\bibinfo {author} {\bibfnamefont {R.}~\bibnamefont
  {Blume-Kohout}}\ and\ \bibinfo {author} {\bibfnamefont {W.~H.}\ \bibnamefont
  {Zurek}},\ }\bibfield  {title} {\bibinfo {title} {Quantum darwinism:
  Entanglement, branches, and the emergent classicality of redundantly stored
  quantum information},\ }\href@noop {} {\bibfield  {journal} {\bibinfo
  {journal} {Physical review A}\ }\textbf {\bibinfo {volume} {73}},\ \bibinfo
  {pages} {062310} (\bibinfo {year} {2006})}\BibitemShut {NoStop}%
\bibitem [{\citenamefont {Korbicz}\ \emph {et~al.}(2014)\citenamefont
  {Korbicz}, \citenamefont {Horodecki},\ and\ \citenamefont
  {Horodecki}}]{korbicz2014objectivity}%
  \BibitemOpen
  \bibfield  {author} {\bibinfo {author} {\bibfnamefont {J.~K.}\ \bibnamefont
  {Korbicz}}, \bibinfo {author} {\bibfnamefont {P.}~\bibnamefont {Horodecki}},\
  and\ \bibinfo {author} {\bibfnamefont {R.}~\bibnamefont {Horodecki}},\
  }\bibfield  {title} {\bibinfo {title} {Objectivity in a noisy photonic
  environment through quantum state information broadcasting},\ }\href@noop {}
  {\bibfield  {journal} {\bibinfo  {journal} {Physical review letters}\
  }\textbf {\bibinfo {volume} {112}},\ \bibinfo {pages} {120402} (\bibinfo
  {year} {2014})}\BibitemShut {NoStop}%
\bibitem [{\citenamefont {Riedel}\ and\ \citenamefont {Zurek}(2010)}]{QD1}%
  \BibitemOpen
  \bibfield  {author} {\bibinfo {author} {\bibfnamefont {C.~J.}\ \bibnamefont
  {Riedel}}\ and\ \bibinfo {author} {\bibfnamefont {W.~H.}\ \bibnamefont
  {Zurek}},\ }\bibfield  {title} {\bibinfo {title} {Quantum darwinism in an
  everyday environment: Huge redundancy in scattered photons},\ }\href
  {https://doi.org/10.1103/PhysRevLett.105.020404} {\bibfield  {journal}
  {\bibinfo  {journal} {Phys. Rev. Lett.}\ }\textbf {\bibinfo {volume} {105}},\
  \bibinfo {pages} {020404} (\bibinfo {year} {2010})}\BibitemShut {NoStop}%
\bibitem [{\citenamefont {Blume-Kohout}\ and\ \citenamefont
  {Zurek}(2008)}]{QD2}%
  \BibitemOpen
  \bibfield  {author} {\bibinfo {author} {\bibfnamefont {R.}~\bibnamefont
  {Blume-Kohout}}\ and\ \bibinfo {author} {\bibfnamefont {W.~H.}\ \bibnamefont
  {Zurek}},\ }\bibfield  {title} {\bibinfo {title} {Quantum darwinism in
  quantum brownian motion},\ }\href
  {https://doi.org/10.1103/PhysRevLett.101.240405} {\bibfield  {journal}
  {\bibinfo  {journal} {Phys. Rev. Lett.}\ }\textbf {\bibinfo {volume} {101}},\
  \bibinfo {pages} {240405} (\bibinfo {year} {2008})}\BibitemShut {NoStop}%
\bibitem [{\citenamefont {Zwolak}\ \emph {et~al.}(2009)\citenamefont {Zwolak},
  \citenamefont {Quan},\ and\ \citenamefont {Zurek}}]{QD3}%
  \BibitemOpen
  \bibfield  {author} {\bibinfo {author} {\bibfnamefont {M.}~\bibnamefont
  {Zwolak}}, \bibinfo {author} {\bibfnamefont {H.~T.}\ \bibnamefont {Quan}},\
  and\ \bibinfo {author} {\bibfnamefont {W.~H.}\ \bibnamefont {Zurek}},\
  }\bibfield  {title} {\bibinfo {title} {Quantum darwinism in a mixed
  environment},\ }\href {https://doi.org/10.1103/PhysRevLett.103.110402}
  {\bibfield  {journal} {\bibinfo  {journal} {Phys. Rev. Lett.}\ }\textbf
  {\bibinfo {volume} {103}},\ \bibinfo {pages} {110402} (\bibinfo {year}
  {2009})}\BibitemShut {NoStop}%
\bibitem [{\citenamefont {Riedel}\ and\ \citenamefont {Zurek}(2011)}]{QD4}%
  \BibitemOpen
  \bibfield  {author} {\bibinfo {author} {\bibfnamefont {C.~J.}\ \bibnamefont
  {Riedel}}\ and\ \bibinfo {author} {\bibfnamefont {W.~H.}\ \bibnamefont
  {Zurek}},\ }\bibfield  {title} {\bibinfo {title} {Redundant information from
  thermal illumination: quantum darwinism in scattered photons},\ }\href
  {https://doi.org/10.1088/1367-2630/13/7/073038} {\bibfield  {journal}
  {\bibinfo  {journal} {New Journal of Physics}\ }\textbf {\bibinfo {volume}
  {13}},\ \bibinfo {pages} {073038} (\bibinfo {year} {2011})}\BibitemShut
  {NoStop}%
\bibitem [{\citenamefont {Horodecki}\ \emph {et~al.}(2015)\citenamefont
  {Horodecki}, \citenamefont {Korbicz},\ and\ \citenamefont
  {Horodecki}}]{horodecki2015quantum}%
  \BibitemOpen
  \bibfield  {author} {\bibinfo {author} {\bibfnamefont {R.}~\bibnamefont
  {Horodecki}}, \bibinfo {author} {\bibfnamefont {J.~K.}\ \bibnamefont
  {Korbicz}},\ and\ \bibinfo {author} {\bibfnamefont {P.}~\bibnamefont
  {Horodecki}},\ }\bibfield  {title} {\bibinfo {title} {Quantum origins of
  objectivity},\ }\href@noop {} {\bibfield  {journal} {\bibinfo  {journal}
  {Physical review A}\ }\textbf {\bibinfo {volume} {91}},\ \bibinfo {pages}
  {032122} (\bibinfo {year} {2015})}\BibitemShut {NoStop}%
\bibitem [{\citenamefont {Brandao}\ \emph {et~al.}(2015)\citenamefont
  {Brandao}, \citenamefont {Piani},\ and\ \citenamefont
  {Horodecki}}]{brandao2015generic}%
  \BibitemOpen
  \bibfield  {author} {\bibinfo {author} {\bibfnamefont {F.~G.}\ \bibnamefont
  {Brandao}}, \bibinfo {author} {\bibfnamefont {M.}~\bibnamefont {Piani}},\
  and\ \bibinfo {author} {\bibfnamefont {P.}~\bibnamefont {Horodecki}},\
  }\bibfield  {title} {\bibinfo {title} {Generic emergence of classical
  features in quantum darwinism},\ }\href@noop {} {\bibfield  {journal}
  {\bibinfo  {journal} {Nature communications}\ }\textbf {\bibinfo {volume}
  {6}},\ \bibinfo {pages} {1} (\bibinfo {year} {2015})}\BibitemShut {NoStop}%
\bibitem [{\citenamefont {Qi}\ and\ \citenamefont
  {Ranard}(2021)}]{qi2021emergent}%
  \BibitemOpen
  \bibfield  {author} {\bibinfo {author} {\bibfnamefont {X.-L.}\ \bibnamefont
  {Qi}}\ and\ \bibinfo {author} {\bibfnamefont {D.}~\bibnamefont {Ranard}},\
  }\bibfield  {title} {\bibinfo {title} {Emergent classicality in general
  multipartite states and channels},\ }\href@noop {} {\bibfield  {journal}
  {\bibinfo  {journal} {Quantum}\ }\textbf {\bibinfo {volume} {5}},\ \bibinfo
  {pages} {555} (\bibinfo {year} {2021})}\BibitemShut {NoStop}%
\bibitem [{\citenamefont {Baldij{\~a}o}\ \emph {et~al.}(2021)\citenamefont
  {Baldij{\~a}o}, \citenamefont {Wagner}, \citenamefont {Duarte}, \citenamefont
  {Amaral},\ and\ \citenamefont {Cunha}}]{baldijao2021emergence}%
  \BibitemOpen
  \bibfield  {author} {\bibinfo {author} {\bibfnamefont {R.}~\bibnamefont
  {Baldij{\~a}o}}, \bibinfo {author} {\bibfnamefont {R.}~\bibnamefont
  {Wagner}}, \bibinfo {author} {\bibfnamefont {C.}~\bibnamefont {Duarte}},
  \bibinfo {author} {\bibfnamefont {B.}~\bibnamefont {Amaral}},\ and\ \bibinfo
  {author} {\bibfnamefont {M.~T.}\ \bibnamefont {Cunha}},\ }\bibfield  {title}
  {\bibinfo {title} {Emergence of noncontextuality under quantum darwinism},\
  }\href@noop {} {\bibfield  {journal} {\bibinfo  {journal} {PRX Quantum}\
  }\textbf {\bibinfo {volume} {2}} (\bibinfo {year} {2021})}\BibitemShut
  {NoStop}%
\bibitem [{\citenamefont {{\c{C}}akmak}\ \emph {et~al.}(2021)\citenamefont
  {{\c{C}}akmak}, \citenamefont {M{\"u}stecapl{\i}o{\u{g}}lu}, \citenamefont
  {Paternostro}, \citenamefont {Vacchini},\ and\ \citenamefont
  {Campbell}}]{ccakmak2021quantum}%
  \BibitemOpen
  \bibfield  {author} {\bibinfo {author} {\bibfnamefont {B.}~\bibnamefont
  {{\c{C}}akmak}}, \bibinfo {author} {\bibfnamefont {{\"O}.~E.}\ \bibnamefont
  {M{\"u}stecapl{\i}o{\u{g}}lu}}, \bibinfo {author} {\bibfnamefont
  {M.}~\bibnamefont {Paternostro}}, \bibinfo {author} {\bibfnamefont
  {B.}~\bibnamefont {Vacchini}},\ and\ \bibinfo {author} {\bibfnamefont
  {S.}~\bibnamefont {Campbell}},\ }\bibfield  {title} {\bibinfo {title}
  {Quantum darwinism in a composite system: Objectivity versus classicality},\
  }\href@noop {} {\bibfield  {journal} {\bibinfo  {journal} {Entropy}\ }\textbf
  {\bibinfo {volume} {23}},\ \bibinfo {pages} {995} (\bibinfo {year}
  {2021})}\BibitemShut {NoStop}%
\bibitem [{\citenamefont {Touil}\ \emph {et~al.}(2022)\citenamefont {Touil},
  \citenamefont {Yan}, \citenamefont {Girolami}, \citenamefont {Deffner},\ and\
  \citenamefont {Zurek}}]{touil2022eavesdropping}%
  \BibitemOpen
  \bibfield  {author} {\bibinfo {author} {\bibfnamefont {A.}~\bibnamefont
  {Touil}}, \bibinfo {author} {\bibfnamefont {B.}~\bibnamefont {Yan}}, \bibinfo
  {author} {\bibfnamefont {D.}~\bibnamefont {Girolami}}, \bibinfo {author}
  {\bibfnamefont {S.}~\bibnamefont {Deffner}},\ and\ \bibinfo {author}
  {\bibfnamefont {W.~H.}\ \bibnamefont {Zurek}},\ }\bibfield  {title} {\bibinfo
  {title} {Eavesdropping on the decohering environment: quantum darwinism,
  amplification, and the origin of objective classical reality},\ }\href@noop
  {} {\bibfield  {journal} {\bibinfo  {journal} {Physical Review Letters}\
  }\textbf {\bibinfo {volume} {128}},\ \bibinfo {pages} {010401} (\bibinfo
  {year} {2022})}\BibitemShut {NoStop}%
\bibitem [{\citenamefont {Clauser}\ \emph {et~al.}(1969)\citenamefont
  {Clauser}, \citenamefont {Horne}, \citenamefont {Shimony},\ and\
  \citenamefont {Holt}}]{clauser1969proposed}%
  \BibitemOpen
  \bibfield  {author} {\bibinfo {author} {\bibfnamefont {J.~F.}\ \bibnamefont
  {Clauser}}, \bibinfo {author} {\bibfnamefont {M.~A.}\ \bibnamefont {Horne}},
  \bibinfo {author} {\bibfnamefont {A.}~\bibnamefont {Shimony}},\ and\ \bibinfo
  {author} {\bibfnamefont {R.~A.}\ \bibnamefont {Holt}},\ }\bibfield  {title}
  {\bibinfo {title} {Proposed experiment to test local hidden-variable
  theories},\ }\href@noop {} {\bibfield  {journal} {\bibinfo  {journal}
  {Physical review letters}\ }\textbf {\bibinfo {volume} {23}},\ \bibinfo
  {pages} {880} (\bibinfo {year} {1969})}\BibitemShut {NoStop}%
\bibitem [{\citenamefont {Brunner}\ \emph {et~al.}(2014)\citenamefont
  {Brunner}, \citenamefont {Cavalcanti}, \citenamefont {Pironio}, \citenamefont
  {Scarani},\ and\ \citenamefont {Wehner}}]{brunner2014bell}%
  \BibitemOpen
  \bibfield  {author} {\bibinfo {author} {\bibfnamefont {N.}~\bibnamefont
  {Brunner}}, \bibinfo {author} {\bibfnamefont {D.}~\bibnamefont {Cavalcanti}},
  \bibinfo {author} {\bibfnamefont {S.}~\bibnamefont {Pironio}}, \bibinfo
  {author} {\bibfnamefont {V.}~\bibnamefont {Scarani}},\ and\ \bibinfo {author}
  {\bibfnamefont {S.}~\bibnamefont {Wehner}},\ }\bibfield  {title} {\bibinfo
  {title} {Bell nonlocality},\ }\href@noop {} {\bibfield  {journal} {\bibinfo
  {journal} {Reviews of Modern Physics}\ }\textbf {\bibinfo {volume} {86}},\
  \bibinfo {pages} {419} (\bibinfo {year} {2014})}\BibitemShut {NoStop}%
\bibitem [{Note1()}]{Note1}%
  \BibitemOpen
  \bibinfo {note} {A possible route to generalization of quantum Darwinism to
  GPTs was proposed in \cite {baldijao2022quantum}. Unfortunately, the authors
  are concerned with defining what an idealized quantum Darwinism process would
  look like in GPTs, more precisely, a general version of a fan-out gate, and
  do not consider the noisy version of such a process.}\BibitemShut {Stop}%
\bibitem [{\citenamefont {Barrett}\ \emph {et~al.}(2006)\citenamefont
  {Barrett}, \citenamefont {Kent},\ and\ \citenamefont
  {Pironio}}]{barrett2006maximally}%
  \BibitemOpen
  \bibfield  {author} {\bibinfo {author} {\bibfnamefont {J.}~\bibnamefont
  {Barrett}}, \bibinfo {author} {\bibfnamefont {A.}~\bibnamefont {Kent}},\ and\
  \bibinfo {author} {\bibfnamefont {S.}~\bibnamefont {Pironio}},\ }\bibfield
  {title} {\bibinfo {title} {Maximally nonlocal and monogamous quantum
  correlations},\ }\href@noop {} {\bibfield  {journal} {\bibinfo  {journal}
  {Physical review letters}\ }\textbf {\bibinfo {volume} {97}},\ \bibinfo
  {pages} {170409} (\bibinfo {year} {2006})}\BibitemShut {NoStop}%
\bibitem [{\citenamefont {Barnum}\ \emph {et~al.}(2007)\citenamefont {Barnum},
  \citenamefont {Barrett}, \citenamefont {Leifer},\ and\ \citenamefont
  {Wilce}}]{barnum2007generalized}%
  \BibitemOpen
  \bibfield  {author} {\bibinfo {author} {\bibfnamefont {H.}~\bibnamefont
  {Barnum}}, \bibinfo {author} {\bibfnamefont {J.}~\bibnamefont {Barrett}},
  \bibinfo {author} {\bibfnamefont {M.}~\bibnamefont {Leifer}},\ and\ \bibinfo
  {author} {\bibfnamefont {A.}~\bibnamefont {Wilce}},\ }\bibfield  {title}
  {\bibinfo {title} {Generalized no-broadcasting theorem},\ }\href@noop {}
  {\bibfield  {journal} {\bibinfo  {journal} {Physical review letters}\
  }\textbf {\bibinfo {volume} {99}},\ \bibinfo {pages} {240501} (\bibinfo
  {year} {2007})}\BibitemShut {NoStop}%
\bibitem [{\citenamefont {Popescu}\ and\ \citenamefont
  {Rohrlich}(1994)}]{popescu1994quantum}%
  \BibitemOpen
  \bibfield  {author} {\bibinfo {author} {\bibfnamefont {S.}~\bibnamefont
  {Popescu}}\ and\ \bibinfo {author} {\bibfnamefont {D.}~\bibnamefont
  {Rohrlich}},\ }\bibfield  {title} {\bibinfo {title} {Quantum nonlocality as
  an axiom},\ }\href@noop {} {\bibfield  {journal} {\bibinfo  {journal}
  {Foundations of Physics}\ }\textbf {\bibinfo {volume} {24}},\ \bibinfo
  {pages} {379} (\bibinfo {year} {1994})}\BibitemShut {NoStop}%
\bibitem [{\citenamefont {M{\"u}ller}\ and\ \citenamefont
  {Masanes}(2016)}]{MM16}%
  \BibitemOpen
  \bibfield  {author} {\bibinfo {author} {\bibfnamefont {M.~P.}\ \bibnamefont
  {M{\"u}ller}}\ and\ \bibinfo {author} {\bibfnamefont {L.}~\bibnamefont
  {Masanes}},\ }\bibinfo {title} {Information-theoretic postulates for quantum
  theory},\ in\ \href {https://doi.org/10.1007/978-94-017-7303-4_5} {\emph
  {\bibinfo {booktitle} {Quantum Theory: Informational Foundations and
  Foils}}},\ \bibinfo {editor} {edited by\ \bibinfo {editor} {\bibfnamefont
  {G.}~\bibnamefont {Chiribella}}\ and\ \bibinfo {editor} {\bibfnamefont
  {R.~W.}\ \bibnamefont {Spekkens}}}\ (\bibinfo  {publisher} {Springer
  Netherlands},\ \bibinfo {address} {Dordrecht},\ \bibinfo {year} {2016})\ pp.\
  \bibinfo {pages} {139--170}\BibitemShut {NoStop}%
\bibitem [{\citenamefont {Pironio}\ \emph {et~al.}(2016)\citenamefont
  {Pironio}, \citenamefont {Scarani},\ and\ \citenamefont
  {Vidick}}]{pironio2016focus}%
  \BibitemOpen
  \bibfield  {author} {\bibinfo {author} {\bibfnamefont {S.}~\bibnamefont
  {Pironio}}, \bibinfo {author} {\bibfnamefont {V.}~\bibnamefont {Scarani}},\
  and\ \bibinfo {author} {\bibfnamefont {T.}~\bibnamefont {Vidick}},\
  }\bibfield  {title} {\bibinfo {title} {Focus on device independent quantum
  information},\ }\href@noop {} {\bibfield  {journal} {\bibinfo  {journal} {New
  Journal of Physics}\ }\textbf {\bibinfo {volume} {18}},\ \bibinfo {pages}
  {100202} (\bibinfo {year} {2016})}\BibitemShut {NoStop}%
\bibitem [{\citenamefont {Ekert}(1991)}]{Ekert1991}%
  \BibitemOpen
  \bibfield  {author} {\bibinfo {author} {\bibfnamefont {A.~K.}\ \bibnamefont
  {Ekert}},\ }\bibfield  {title} {\bibinfo {title} {Quantum cryptography based
  on bell's theorem},\ }\href {https://doi.org/10.1103/PhysRevLett.67.661}
  {\bibfield  {journal} {\bibinfo  {journal} {Phys. Rev. Lett.}\ }\textbf
  {\bibinfo {volume} {67}},\ \bibinfo {pages} {661} (\bibinfo {year}
  {1991})}\BibitemShut {NoStop}%
\bibitem [{\citenamefont {Pironio}\ \emph {et~al.}(2010)\citenamefont
  {Pironio}, \citenamefont {Ac{\'\i}n}, \citenamefont {Massar}, \citenamefont
  {de~La~Giroday}, \citenamefont {Matsukevich}, \citenamefont {Maunz},
  \citenamefont {Olmschenk}, \citenamefont {Hayes}, \citenamefont {Luo},
  \citenamefont {Manning} \emph {et~al.}}]{pironio2010random}%
  \BibitemOpen
  \bibfield  {author} {\bibinfo {author} {\bibfnamefont {S.}~\bibnamefont
  {Pironio}}, \bibinfo {author} {\bibfnamefont {A.}~\bibnamefont {Ac{\'\i}n}},
  \bibinfo {author} {\bibfnamefont {S.}~\bibnamefont {Massar}}, \bibinfo
  {author} {\bibfnamefont {A.~B.}\ \bibnamefont {de~La~Giroday}}, \bibinfo
  {author} {\bibfnamefont {D.~N.}\ \bibnamefont {Matsukevich}}, \bibinfo
  {author} {\bibfnamefont {P.}~\bibnamefont {Maunz}}, \bibinfo {author}
  {\bibfnamefont {S.}~\bibnamefont {Olmschenk}}, \bibinfo {author}
  {\bibfnamefont {D.}~\bibnamefont {Hayes}}, \bibinfo {author} {\bibfnamefont
  {L.}~\bibnamefont {Luo}}, \bibinfo {author} {\bibfnamefont {T.~A.}\
  \bibnamefont {Manning}}, \emph {et~al.},\ }\bibfield  {title} {\bibinfo
  {title} {Random numbers certified by bell’s theorem},\ }\href@noop {}
  {\bibfield  {journal} {\bibinfo  {journal} {Nature}\ }\textbf {\bibinfo
  {volume} {464}},\ \bibinfo {pages} {1021} (\bibinfo {year}
  {2010})}\BibitemShut {NoStop}%
\bibitem [{\citenamefont {Cabello}\ \emph {et~al.}(2011)\citenamefont
  {Cabello}, \citenamefont {D'Ambrosio}, \citenamefont {Nagali},\ and\
  \citenamefont {Sciarrino}}]{CabelloEtAl11}%
  \BibitemOpen
  \bibfield  {author} {\bibinfo {author} {\bibfnamefont {A.}~\bibnamefont
  {Cabello}}, \bibinfo {author} {\bibfnamefont {V.}~\bibnamefont {D'Ambrosio}},
  \bibinfo {author} {\bibfnamefont {E.}~\bibnamefont {Nagali}},\ and\ \bibinfo
  {author} {\bibfnamefont {F.}~\bibnamefont {Sciarrino}},\ }\bibfield  {title}
  {\bibinfo {title} {Hybrid ququart-encoded quantum cryptography protected by
  kochen-specker contextuality},\ }\href
  {https://doi.org/10.1103/PhysRevA.84.030302} {\bibfield  {journal} {\bibinfo
  {journal} {Phys. Rev. A}\ }\textbf {\bibinfo {volume} {84}},\ \bibinfo
  {pages} {030302} (\bibinfo {year} {2011})}\BibitemShut {NoStop}%
\bibitem [{\citenamefont {Hall}(2016)}]{hall2016significance}%
  \BibitemOpen
  \bibfield  {author} {\bibinfo {author} {\bibfnamefont {M.~J.}\ \bibnamefont
  {Hall}},\ }\bibfield  {title} {\bibinfo {title} {The significance of
  measurement independence for bell inequalities and locality},\ }in\
  \href@noop {} {\emph {\bibinfo {booktitle} {At the frontier of spacetime}}}\
  (\bibinfo  {publisher} {Springer},\ \bibinfo {year} {2016})\ pp.\ \bibinfo
  {pages} {189--204}\BibitemShut {NoStop}%
\bibitem [{\citenamefont {Chaves}\ \emph {et~al.}(2021)\citenamefont {Chaves},
  \citenamefont {Moreno}, \citenamefont {Polino}, \citenamefont {Poderini},
  \citenamefont {Agresti}, \citenamefont {Suprano}, \citenamefont {Barros},
  \citenamefont {Carvacho}, \citenamefont {Wolfe}, \citenamefont {Canabarro}
  \emph {et~al.}}]{chaves2021causal}%
  \BibitemOpen
  \bibfield  {author} {\bibinfo {author} {\bibfnamefont {R.}~\bibnamefont
  {Chaves}}, \bibinfo {author} {\bibfnamefont {G.}~\bibnamefont {Moreno}},
  \bibinfo {author} {\bibfnamefont {E.}~\bibnamefont {Polino}}, \bibinfo
  {author} {\bibfnamefont {D.}~\bibnamefont {Poderini}}, \bibinfo {author}
  {\bibfnamefont {I.}~\bibnamefont {Agresti}}, \bibinfo {author} {\bibfnamefont
  {A.}~\bibnamefont {Suprano}}, \bibinfo {author} {\bibfnamefont {M.~R.}\
  \bibnamefont {Barros}}, \bibinfo {author} {\bibfnamefont {G.}~\bibnamefont
  {Carvacho}}, \bibinfo {author} {\bibfnamefont {E.}~\bibnamefont {Wolfe}},
  \bibinfo {author} {\bibfnamefont {A.}~\bibnamefont {Canabarro}}, \emph
  {et~al.},\ }\bibfield  {title} {\bibinfo {title} {Causal networks and freedom
  of choice in bell’s theorem},\ }\href@noop {} {\bibfield  {journal}
  {\bibinfo  {journal} {PRX Quantum}\ }\textbf {\bibinfo {volume} {2}},\
  \bibinfo {pages} {040323} (\bibinfo {year} {2021})}\BibitemShut {NoStop}%
\bibitem [{\citenamefont {{\v{S}}upi{\'{c}}}\ and\ \citenamefont
  {Bowles}(2020)}]{Supic2020}%
  \BibitemOpen
  \bibfield  {author} {\bibinfo {author} {\bibfnamefont {I.}~\bibnamefont
  {{\v{S}}upi{\'{c}}}}\ and\ \bibinfo {author} {\bibfnamefont {J.}~\bibnamefont
  {Bowles}},\ }\bibfield  {title} {\bibinfo {title} {Self-testing of quantum
  systems: a review},\ }\href {https://doi.org/10.22331/q-2020-09-30-337}
  {\bibfield  {journal} {\bibinfo  {journal} {{Quantum}}\ }\textbf {\bibinfo
  {volume} {4}},\ \bibinfo {pages} {337} (\bibinfo {year} {2020})}\BibitemShut
  {NoStop}%
\bibitem [{\citenamefont {Masanes}\ \emph {et~al.}(2006)\citenamefont
  {Masanes}, \citenamefont {Acin},\ and\ \citenamefont {Gisin}}]{Masanes2006}%
  \BibitemOpen
  \bibfield  {author} {\bibinfo {author} {\bibfnamefont {L.}~\bibnamefont
  {Masanes}}, \bibinfo {author} {\bibfnamefont {A.}~\bibnamefont {Acin}},\ and\
  \bibinfo {author} {\bibfnamefont {N.}~\bibnamefont {Gisin}},\ }\bibfield
  {title} {\bibinfo {title} {General properties of nonsignaling theories},\
  }\href {https://doi.org/10.1103/PhysRevA.73.012112} {\bibfield  {journal}
  {\bibinfo  {journal} {Phys. Rev. A}\ }\textbf {\bibinfo {volume} {73}},\
  \bibinfo {pages} {012112} (\bibinfo {year} {2006})}\BibitemShut {NoStop}%
\bibitem [{\citenamefont {Dhara}\ \emph {et~al.}(2013)\citenamefont {Dhara},
  \citenamefont {Prettico},\ and\ \citenamefont {Ac\'{\i}n}}]{Dhara2013}%
  \BibitemOpen
  \bibfield  {author} {\bibinfo {author} {\bibfnamefont {C.}~\bibnamefont
  {Dhara}}, \bibinfo {author} {\bibfnamefont {G.}~\bibnamefont {Prettico}},\
  and\ \bibinfo {author} {\bibfnamefont {A.}~\bibnamefont {Ac\'{\i}n}},\
  }\bibfield  {title} {\bibinfo {title} {Maximal quantum randomness in bell
  tests},\ }\href {https://doi.org/10.1103/PhysRevA.88.052116} {\bibfield
  {journal} {\bibinfo  {journal} {Phys. Rev. A}\ }\textbf {\bibinfo {volume}
  {88}},\ \bibinfo {pages} {052116} (\bibinfo {year} {2013})}\BibitemShut
  {NoStop}%
\bibitem [{\citenamefont {Navascu\'es}\ \emph {et~al.}(2007)\citenamefont
  {Navascu\'es}, \citenamefont {Pironio},\ and\ \citenamefont
  {Ac\'{\i}n}}]{NPA_2007}%
  \BibitemOpen
  \bibfield  {author} {\bibinfo {author} {\bibfnamefont {M.}~\bibnamefont
  {Navascu\'es}}, \bibinfo {author} {\bibfnamefont {S.}~\bibnamefont
  {Pironio}},\ and\ \bibinfo {author} {\bibfnamefont {A.}~\bibnamefont
  {Ac\'{\i}n}},\ }\bibfield  {title} {\bibinfo {title} {Bounding the set of
  quantum correlations},\ }\href
  {https://doi.org/10.1103/PhysRevLett.98.010401} {\bibfield  {journal}
  {\bibinfo  {journal} {Phys. Rev. Lett.}\ }\textbf {\bibinfo {volume} {98}},\
  \bibinfo {pages} {010401} (\bibinfo {year} {2007})}\BibitemShut {NoStop}%
\bibitem [{\citenamefont {Wang}\ \emph {et~al.}(2016)\citenamefont {Wang},
  \citenamefont {Wu},\ and\ \citenamefont {Scarani}}]{Wang2016}%
  \BibitemOpen
  \bibfield  {author} {\bibinfo {author} {\bibfnamefont {Y.}~\bibnamefont
  {Wang}}, \bibinfo {author} {\bibfnamefont {X.}~\bibnamefont {Wu}},\ and\
  \bibinfo {author} {\bibfnamefont {V.}~\bibnamefont {Scarani}},\ }\bibfield
  {title} {\bibinfo {title} {All the self-testings of the singlet for two
  binary measurements},\ }\href {https://doi.org/10.1088/1367-2630/18/2/025021}
  {\bibfield  {journal} {\bibinfo  {journal} {New Journal of Physics}\ }\textbf
  {\bibinfo {volume} {18}},\ \bibinfo {pages} {025021} (\bibinfo {year}
  {2016})}\BibitemShut {NoStop}%
\bibitem [{\citenamefont {Preskill}(1998)}]{preskill1998lecture}%
  \BibitemOpen
  \bibfield  {author} {\bibinfo {author} {\bibfnamefont {J.}~\bibnamefont
  {Preskill}},\ }\bibfield  {title} {\bibinfo {title} {Lecture notes for
  physics 229: Quantum information and computation},\ }\href@noop {} {\bibfield
   {journal} {\bibinfo  {journal} {California Institute of Technology}\
  }\textbf {\bibinfo {volume} {16}},\ \bibinfo {pages} {1} (\bibinfo {year}
  {1998})}\BibitemShut {NoStop}%
\bibitem [{\citenamefont {Fedrizzi}\ \emph {et~al.}(2007)\citenamefont
  {Fedrizzi}, \citenamefont {Herbst}, \citenamefont {Poppe}, \citenamefont
  {Jennewein},\ and\ \citenamefont {Zeilinger}}]{fedrizzi2007wavelength}%
  \BibitemOpen
  \bibfield  {author} {\bibinfo {author} {\bibfnamefont {A.}~\bibnamefont
  {Fedrizzi}}, \bibinfo {author} {\bibfnamefont {T.}~\bibnamefont {Herbst}},
  \bibinfo {author} {\bibfnamefont {A.}~\bibnamefont {Poppe}}, \bibinfo
  {author} {\bibfnamefont {T.}~\bibnamefont {Jennewein}},\ and\ \bibinfo
  {author} {\bibfnamefont {A.}~\bibnamefont {Zeilinger}},\ }\bibfield  {title}
  {\bibinfo {title} {A wavelength-tunable fiber-coupled source of narrowband
  entangled photons},\ }\href@noop {} {\bibfield  {journal} {\bibinfo
  {journal} {Optics Express}\ }\textbf {\bibinfo {volume} {15}},\ \bibinfo
  {pages} {15377} (\bibinfo {year} {2007})}\BibitemShut {NoStop}%
\bibitem [{\citenamefont {James}\ \emph {et~al.}(2005)\citenamefont {James},
  \citenamefont {Kwiat}, \citenamefont {Munro},\ and\ \citenamefont
  {White}}]{james2005measurement}%
  \BibitemOpen
  \bibfield  {author} {\bibinfo {author} {\bibfnamefont {D.~F.}\ \bibnamefont
  {James}}, \bibinfo {author} {\bibfnamefont {P.~G.}\ \bibnamefont {Kwiat}},
  \bibinfo {author} {\bibfnamefont {W.~J.}\ \bibnamefont {Munro}},\ and\
  \bibinfo {author} {\bibfnamefont {A.~G.}\ \bibnamefont {White}},\ }\bibfield
  {title} {\bibinfo {title} {On the measurement of qubits},\ }in\ \href@noop {}
  {\emph {\bibinfo {booktitle} {Asymptotic Theory of Quantum Statistical
  Inference: Selected Papers}}}\ (\bibinfo  {publisher} {World Scientific},\
  \bibinfo {year} {2005})\ pp.\ \bibinfo {pages} {509--538}\BibitemShut
  {NoStop}%
\bibitem [{\citenamefont {Osborne}\ and\ \citenamefont
  {Verstraete}(2006)}]{osborne2006general}%
  \BibitemOpen
  \bibfield  {author} {\bibinfo {author} {\bibfnamefont {T.~J.}\ \bibnamefont
  {Osborne}}\ and\ \bibinfo {author} {\bibfnamefont {F.}~\bibnamefont
  {Verstraete}},\ }\bibfield  {title} {\bibinfo {title} {General monogamy
  inequality for bipartite qubit entanglement},\ }\href@noop {} {\bibfield
  {journal} {\bibinfo  {journal} {Physical review letters}\ }\textbf {\bibinfo
  {volume} {96}},\ \bibinfo {pages} {220503} (\bibinfo {year}
  {2006})}\BibitemShut {NoStop}%
\bibitem [{\citenamefont {Poderini}\ \emph {et~al.}(2022)\citenamefont
  {Poderini}, \citenamefont {Polino}, \citenamefont {Rodari}, \citenamefont
  {Suprano}, \citenamefont {Chaves},\ and\ \citenamefont
  {Sciarrino}}]{poderini2022ab}%
  \BibitemOpen
  \bibfield  {author} {\bibinfo {author} {\bibfnamefont {D.}~\bibnamefont
  {Poderini}}, \bibinfo {author} {\bibfnamefont {E.}~\bibnamefont {Polino}},
  \bibinfo {author} {\bibfnamefont {G.}~\bibnamefont {Rodari}}, \bibinfo
  {author} {\bibfnamefont {A.}~\bibnamefont {Suprano}}, \bibinfo {author}
  {\bibfnamefont {R.}~\bibnamefont {Chaves}},\ and\ \bibinfo {author}
  {\bibfnamefont {F.}~\bibnamefont {Sciarrino}},\ }\bibfield  {title} {\bibinfo
  {title} {Ab initio experimental violation of bell inequalities},\ }\href@noop
  {} {\bibfield  {journal} {\bibinfo  {journal} {Physical Review Research}\
  }\textbf {\bibinfo {volume} {4}},\ \bibinfo {pages} {013159} (\bibinfo {year}
  {2022})}\BibitemShut {NoStop}%
\bibitem [{\citenamefont {Girolami}\ \emph {et~al.}(2022)\citenamefont
  {Girolami}, \citenamefont {Touil}, \citenamefont {Yan}, \citenamefont
  {Deffner},\ and\ \citenamefont {Zurek}}]{girolami2022redundantly}%
  \BibitemOpen
  \bibfield  {author} {\bibinfo {author} {\bibfnamefont {D.}~\bibnamefont
  {Girolami}}, \bibinfo {author} {\bibfnamefont {A.}~\bibnamefont {Touil}},
  \bibinfo {author} {\bibfnamefont {B.}~\bibnamefont {Yan}}, \bibinfo {author}
  {\bibfnamefont {S.}~\bibnamefont {Deffner}},\ and\ \bibinfo {author}
  {\bibfnamefont {W.~H.}\ \bibnamefont {Zurek}},\ }\bibfield  {title} {\bibinfo
  {title} {Redundantly amplified information suppresses quantum correlations in
  many-body systems},\ }\href@noop {} {\bibfield  {journal} {\bibinfo
  {journal} {arXiv preprint arXiv:2202.09328}\ } (\bibinfo {year}
  {2022})}\BibitemShut {NoStop}%
\bibitem [{\citenamefont {Ciampini}\ \emph {et~al.}(2018)\citenamefont
  {Ciampini}, \citenamefont {Pinna}, \citenamefont {Mataloni},\ and\
  \citenamefont {Paternostro}}]{ciampini2018experimental}%
  \BibitemOpen
  \bibfield  {author} {\bibinfo {author} {\bibfnamefont {M.~A.}\ \bibnamefont
  {Ciampini}}, \bibinfo {author} {\bibfnamefont {G.}~\bibnamefont {Pinna}},
  \bibinfo {author} {\bibfnamefont {P.}~\bibnamefont {Mataloni}},\ and\
  \bibinfo {author} {\bibfnamefont {M.}~\bibnamefont {Paternostro}},\
  }\bibfield  {title} {\bibinfo {title} {Experimental signature of quantum
  darwinism in photonic cluster states},\ }\href@noop {} {\bibfield  {journal}
  {\bibinfo  {journal} {Physical Review A}\ }\textbf {\bibinfo {volume} {98}},\
  \bibinfo {pages} {020101} (\bibinfo {year} {2018})}\BibitemShut {NoStop}%
\bibitem [{\citenamefont {Chen}\ \emph {et~al.}(2019)\citenamefont {Chen},
  \citenamefont {Zhong}, \citenamefont {Li}, \citenamefont {Wu}, \citenamefont
  {Wang}, \citenamefont {Li}, \citenamefont {Liu}, \citenamefont {Lu},\ and\
  \citenamefont {Pan}}]{chen2019emergence}%
  \BibitemOpen
  \bibfield  {author} {\bibinfo {author} {\bibfnamefont {M.-C.}\ \bibnamefont
  {Chen}}, \bibinfo {author} {\bibfnamefont {H.-S.}\ \bibnamefont {Zhong}},
  \bibinfo {author} {\bibfnamefont {Y.}~\bibnamefont {Li}}, \bibinfo {author}
  {\bibfnamefont {D.}~\bibnamefont {Wu}}, \bibinfo {author} {\bibfnamefont
  {X.-L.}\ \bibnamefont {Wang}}, \bibinfo {author} {\bibfnamefont
  {L.}~\bibnamefont {Li}}, \bibinfo {author} {\bibfnamefont {N.-L.}\
  \bibnamefont {Liu}}, \bibinfo {author} {\bibfnamefont {C.-Y.}\ \bibnamefont
  {Lu}},\ and\ \bibinfo {author} {\bibfnamefont {J.-W.}\ \bibnamefont {Pan}},\
  }\bibfield  {title} {\bibinfo {title} {Emergence of classical objectivity of
  quantum darwinism in a photonic quantum simulator},\ }\href@noop {}
  {\bibfield  {journal} {\bibinfo  {journal} {Science Bulletin}\ }\textbf
  {\bibinfo {volume} {64}},\ \bibinfo {pages} {580} (\bibinfo {year}
  {2019})}\BibitemShut {NoStop}%
\bibitem [{\citenamefont {Unden}\ \emph {et~al.}(2019)\citenamefont {Unden},
  \citenamefont {Louzon}, \citenamefont {Zwolak}, \citenamefont {Zurek},\ and\
  \citenamefont {Jelezko}}]{unden2019}%
  \BibitemOpen
  \bibfield  {author} {\bibinfo {author} {\bibfnamefont {T.~K.}\ \bibnamefont
  {Unden}}, \bibinfo {author} {\bibfnamefont {D.}~\bibnamefont {Louzon}},
  \bibinfo {author} {\bibfnamefont {M.}~\bibnamefont {Zwolak}}, \bibinfo
  {author} {\bibfnamefont {W.~H.}\ \bibnamefont {Zurek}},\ and\ \bibinfo
  {author} {\bibfnamefont {F.}~\bibnamefont {Jelezko}},\ }\bibfield  {title}
  {\bibinfo {title} {Revealing the emergence of classicality using
  nitrogen-vacancy centers},\ }\href
  {https://doi.org/10.1103/PhysRevLett.123.140402} {\bibfield  {journal}
  {\bibinfo  {journal} {Phys. Rev. Lett.}\ }\textbf {\bibinfo {volume} {123}},\
  \bibinfo {pages} {140402} (\bibinfo {year} {2019})}\BibitemShut {NoStop}%
\bibitem [{\citenamefont {Bong}\ \emph {et~al.}(2020)\citenamefont {Bong},
  \citenamefont {Utreras-Alarc{\'o}n}, \citenamefont {Ghafari}, \citenamefont
  {Liang}, \citenamefont {Tischler}, \citenamefont {Cavalcanti}, \citenamefont
  {Pryde},\ and\ \citenamefont {Wiseman}}]{bong2020strong}%
  \BibitemOpen
  \bibfield  {author} {\bibinfo {author} {\bibfnamefont {K.-W.}\ \bibnamefont
  {Bong}}, \bibinfo {author} {\bibfnamefont {A.}~\bibnamefont
  {Utreras-Alarc{\'o}n}}, \bibinfo {author} {\bibfnamefont {F.}~\bibnamefont
  {Ghafari}}, \bibinfo {author} {\bibfnamefont {Y.-C.}\ \bibnamefont {Liang}},
  \bibinfo {author} {\bibfnamefont {N.}~\bibnamefont {Tischler}}, \bibinfo
  {author} {\bibfnamefont {E.~G.}\ \bibnamefont {Cavalcanti}}, \bibinfo
  {author} {\bibfnamefont {G.~J.}\ \bibnamefont {Pryde}},\ and\ \bibinfo
  {author} {\bibfnamefont {H.~M.}\ \bibnamefont {Wiseman}},\ }\bibfield
  {title} {\bibinfo {title} {A strong no-go theorem on the wigner’s friend
  paradox},\ }\href@noop {} {\bibfield  {journal} {\bibinfo  {journal} {Nature
  Physics}\ }\textbf {\bibinfo {volume} {16}},\ \bibinfo {pages} {1199}
  (\bibinfo {year} {2020})}\BibitemShut {NoStop}%
\bibitem [{\citenamefont {Moreno}\ \emph {et~al.}(2021)\citenamefont {Moreno},
  \citenamefont {Nery}, \citenamefont {Duarte},\ and\ \citenamefont
  {Chaves}}]{Moreno2021}%
  \BibitemOpen
  \bibfield  {author} {\bibinfo {author} {\bibfnamefont {G.}~\bibnamefont
  {Moreno}}, \bibinfo {author} {\bibfnamefont {R.}~\bibnamefont {Nery}},
  \bibinfo {author} {\bibfnamefont {C.}~\bibnamefont {Duarte}},\ and\ \bibinfo
  {author} {\bibfnamefont {R.}~\bibnamefont {Chaves}},\ }\href
  {https://doi.org/10.48550/ARXIV.2112.11223} {\bibinfo {title} {Events in
  quantum mechanics are maximally non-absolute}} (\bibinfo {year}
  {2021})\BibitemShut {NoStop}%
\bibitem [{\citenamefont {Wiseman}\ \emph {et~al.}(2022)\citenamefont
  {Wiseman}, \citenamefont {Cavalcanti},\ and\ \citenamefont
  {Rieffel}}]{wiseman2022thoughtful}%
  \BibitemOpen
  \bibfield  {author} {\bibinfo {author} {\bibfnamefont {H.~M.}\ \bibnamefont
  {Wiseman}}, \bibinfo {author} {\bibfnamefont {E.~G.}\ \bibnamefont
  {Cavalcanti}},\ and\ \bibinfo {author} {\bibfnamefont {E.~G.}\ \bibnamefont
  {Rieffel}},\ }\bibfield  {title} {\bibinfo {title} {A" thoughtful" local
  friendliness no-go theorem: a prospective experiment with new assumptions to
  suit},\ }\href@noop {} {\bibfield  {journal} {\bibinfo  {journal} {arXiv
  preprint arXiv:2209.08491}\ } (\bibinfo {year} {2022})}\BibitemShut {NoStop}%
\bibitem [{\citenamefont {Baldij{\~a}o}\ \emph {et~al.}(2022)\citenamefont
  {Baldij{\~a}o}, \citenamefont {Krumm}, \citenamefont {Garner},\ and\
  \citenamefont {M{\"u}ller}}]{baldijao2022quantum}%
  \BibitemOpen
  \bibfield  {author} {\bibinfo {author} {\bibfnamefont {R.~D.}\ \bibnamefont
  {Baldij{\~a}o}}, \bibinfo {author} {\bibfnamefont {M.}~\bibnamefont {Krumm}},
  \bibinfo {author} {\bibfnamefont {A.~J.}\ \bibnamefont {Garner}},\ and\
  \bibinfo {author} {\bibfnamefont {M.~P.}\ \bibnamefont {M{\"u}ller}},\
  }\bibfield  {title} {\bibinfo {title} {Quantum darwinism and the spreading of
  classical information in non-classical theories},\ }\href@noop {} {\bibfield
  {journal} {\bibinfo  {journal} {Quantum}\ }\textbf {\bibinfo {volume} {6}},\
  \bibinfo {pages} {636} (\bibinfo {year} {2022})}\BibitemShut {NoStop}%
\bibitem [{\citenamefont {Navascu{\'{e}}s}\ \emph {et~al.}(2008)\citenamefont
  {Navascu{\'{e}}s}, \citenamefont {Pironio},\ and\ \citenamefont
  {Ac{\'{\i}}n}}]{NPA_2008}%
  \BibitemOpen
  \bibfield  {author} {\bibinfo {author} {\bibfnamefont {M.}~\bibnamefont
  {Navascu{\'{e}}s}}, \bibinfo {author} {\bibfnamefont {S.}~\bibnamefont
  {Pironio}},\ and\ \bibinfo {author} {\bibfnamefont {A.}~\bibnamefont
  {Ac{\'{\i}}n}},\ }\bibfield  {title} {\bibinfo {title} {A convergent
  hierarchy of semidefinite programs characterizing the set of quantum
  correlations},\ }\href {https://doi.org/10.1088/1367-2630/10/7/073013}
  {\bibfield  {journal} {\bibinfo  {journal} {New Journal of Physics}\ }\textbf
  {\bibinfo {volume} {10}},\ \bibinfo {pages} {073013} (\bibinfo {year}
  {2008})}\BibitemShut {NoStop}%
\bibitem [{\citenamefont {Boyd}(2020)}]{boyd2020nonlinear}%
  \BibitemOpen
  \bibfield  {author} {\bibinfo {author} {\bibfnamefont {R.~W.}\ \bibnamefont
  {Boyd}},\ }\href@noop {} {\emph {\bibinfo {title} {Nonlinear optics}}}\
  (\bibinfo  {publisher} {Academic press},\ \bibinfo {year} {2020})\BibitemShut
  {NoStop}%
\bibitem [{\citenamefont {Chang}(2012)}]{chang2012stochastic}%
  \BibitemOpen
  \bibfield  {author} {\bibinfo {author} {\bibfnamefont {K.-H.}\ \bibnamefont
  {Chang}},\ }\bibfield  {title} {\bibinfo {title} {Stochastic nelder--mead
  simplex method--a new globally convergent direct search method for simulation
  optimization},\ }\href@noop {} {\bibfield  {journal} {\bibinfo  {journal}
  {European journal of operational research}\ }\textbf {\bibinfo {volume}
  {220}},\ \bibinfo {pages} {684} (\bibinfo {year} {2012})}\BibitemShut
  {NoStop}%
\end{thebibliography}%

\begin{widetext}

\section{Appendix}

\subsection{Proof of Result 1}
\label{sec:app_proof1}
\begin{proof}
    We start by summing all the conditions $p(b_k = a, a|x_k^*) \ge 1-\delta$ and expanding them in terms of the joint distribution
    $p(b_1=a, b_2=a, \ldots, b_n=a, a|x_1^*,\ldots,x_n^*), p(b_1\neq a, b_2=a, \ldots, b_n=a, a|x_1^*,\ldots,x_n^*), \ldots$.
    To simplify the notation we use $p_s = p_{\beta_1,\beta_2,\ldots,\beta_n}$, where $s$ is a string of bits $\beta_i \in \{0,1\}$, denoting the term in which $b_i = a$ or $b_i \neq a$, respectively with $\beta_i = 0$ and $\beta_i = 1$.
    So for example $p_0 = p_{00\ldots 0}$ represents the term where all the $b_i$ agree with $a$, while $p_1 = p_{10\ldots 0}$ represents the one where only $b_1$ does not.
    Calling $H(s)$ the hamming distance between $00\ldots 0$ and $s$, i.e. the number of $1$s in the bitstring $s$, we obtain:
    \begin{align}
        n(1-\delta) &\le \sum_k p(b_k = a|x_k^*) =
        \sum_{s \in \{0,1\}^n} (n - H(s)) p_s
        \le \\
        &\le p_0 + (n-1) \left( p_0 + \sum_{s: H(s) \ge 1} p_s \right) \le p_0 + (n-1)
    \end{align}
    where we used the normalization condition for the $p_s$.
    From this we directly obtain \eqref{eq:bound_oout_friends}.
\end{proof}

\subsubsection*{Thightness of bound}
\begin{proof}
 To prove the tightness of the bound consider the distribution
    $p(a, b_1, \ldots, b_n | x_1, \ldots, x_n)$ defined as follows by
    highlighting the settings and the variables associated with special settings $x^*_i$: 
    \begin{equation}
        p(a, b_1, \ldots, b_n | x^*_{i_1}, \ldots, x^*_{i_k}, x_{i_{k+1}}, \ldots, x_{i_n}) = 
        p_\delta(a, b_{i_1}, \ldots, b_{i_k}) p_U(b_{i_{k+1}}, \ldots, b_{i_n})
        \label{eq:distr_fact}
    \end{equation}
    where
    \begin{equation}
        p_\delta(a, b_{i_1}, \ldots, b_{i_k}) = \left\{
        \begin{aligned}
            &(1 - k\delta)/2 & \text{for}\quad a=b_{i_l} \forall l \le k\\
            &\delta/2 & \text{for}\quad a=b_{i_l} \forall l \le k, l\neq m, b_{i_m} \neq a\\
            &0 &\text{elsewhere}
        \end{aligned}
        \right.
        \label{eq:tight_distr}
    \end{equation}
    while $p_U(b_{i_{k+1}}, \ldots, b_{i_n}) = 2^{k-n}$ is the uniform distribution over the remaining variables.
    First, notice that this distribution for $n+1$ variables reduces to the one for $n$ variables by marginalizing on any $b_i$.
    This is obvious when $x_{n+1} \neq x^*_{n+1}$, because of the definition \eqref{eq:distr_fact}, while for $x_{n+1} = x^*_{n+1}$ we have
    \begin{align}
        &p(a,b_1,\ldots,b_n | x^*_{i_1}, \ldots, x^*_{i_k}) = p(a,b_1,\ldots,b_n, b_{n+1} = a | x^*_{i_1}, \ldots, x^*_{i_k}, x^*_{n+1}) +\\ & + p(a,b_1,\ldots,b_n, b_{n+1} \neq a | x^*_{i_1}, \ldots, x^*_{i_k}, x^*_{n+1}) = (1-k\delta)/2 \quad \text{when}\quad a=b_{i_l} \forall l \le k \\
        &p(a,b_1,\ldots,b_n | x^*_{i_1}, \ldots, x^*_{i_k}) = p(a,b_1,\ldots,b_n, b_{n+1} = a| x^*_{i_1}, \ldots, x^*_{i_k}, x^*_{n+1}) + \\
        & + p(a,b_1,\ldots,b_n, b_{n+1} \neq a| x^*_{i_1}, \ldots, x^*_{i_k}, x^*_{n+1}) = \delta/2 \quad \text{when}\quad a=b_{i_l} \forall l \le k, l\neq m, b_{i_m} \neq a
    \end{align}
    It is now easy to show that these distributions satisfy \eqref{eq:objcondition}. Indeed the condition is trivially valid for the case $n=1$, while for a general one, \eqref{eq:objcondition} applies to the marginal on all but one of the $b_i$, which corresponds to the distribution \eqref{eq:distr_fact} for $n=1$ as we just proved.
    
    Similarly to show that \eqref{eq:tight_distr} satisfies the no-signaling condition $\sum_{a, b_j} p(a,b_i,b_j | x_i, x_j) = \sum_{a, b_j} p(a,b_i,b_j | x_i, x_j') \quad \forall \; x_j, x_j'$, we again use the fact that marginalizing on $b_j$ gives \eqref{eq:distr_fact} for $n=1$, which is independent of the choice of $x_j$.
\end{proof}

\subsection{Proof of Result 2}
\begin{proof}
The proof follows closely Ref. \cite{Moreno2021}. Given a distribution $p(b_1,b_2|x_1,x_2,a)$, let us define the following quantities:
\begin{eqnarray}
\gamma^{(a)}_{x_1,x_2} \equiv p(b_1=0,b_2 = 0|x_1,x_2,a),\\
\beta^{(a)}_{x_1} \equiv p(b_1=0|x_1,a),\\
\eta^{(a)}_{x_2} \equiv p(b_2=0|x_2,a).
\end{eqnarray}
Now, using the definition $\langle B^{x_1}_1B^{x_2}_2\rangle_a = \sum_{}(-1)^{b_1+b_2}p(b_1,b_2\vert x_1,x_2,a)$, the NS condition, and the fact that $\max\{0,\beta^{(a)}_{x_1} + \eta^{(a)}_{x_2} - 1\}\leq \gamma^{(a)}_{x_1,x_2}\leq \min\{\beta^{(a)}_{x_1},\eta^{(a)}_{x_2}\}$, one can show that
\begin{align}
2\left|\beta^{(a)}_{x_1} + \eta^{(a)}_{x_2} - 1\right| - 1 \leq \langle B^{x_1}_1B^{x_2}_2\rangle_{a}\leq 1 - 2\left|\beta^{(a)}_{x_1} - \eta^{(a)}_{x_2}\right|.
\end{align}
which can be used to bound the quantity $\CHSH_{\delta,\epsilon}^{(a)} = \langle B^0_1B^0_2\rangle_a +\langle B^0_1B^1_2\rangle_a - \langle B^1_1B^0_2\rangle_a + \langle B^1_1B^1_2\rangle_a$, as
\begin{eqnarray}
\CHSH_{\delta,\epsilon}^{(a)}\leq \langle B^0_1B^0_2\rangle_a + 3 - 2J^{(a)}_{\delta},
\end{eqnarray}
where

\begin{align}
\nonumber
J^{(a)}_{\delta} & \equiv \left|\beta^{(a)}_{0} - \eta^{(a)}_{1}\right| + \left|-\beta^{(a)}_{1} + \eta^{(a)}_{1}\right| + \left|\beta^{(a)}_{1} + \eta^{(a)}_{0} - 1\right|\\
\nonumber
& \geq \left|\beta_0^{(a)} - \beta_1^{(a)}\right| + \left|\beta^{(a)}_{1} + \eta^{(a)}_{0} - 1\right|\\
\nonumber
& \geq \left|\beta_0^{(a)} + \eta_0^{(a)} - 1\right| = \left|p(b_1=0|x_1=0,a) + p(b_2=0|x_2=0,a) - 1\right|.
\end{align}
All the inequalities above are obtained via a direct application of the triangle inequality. In the second line above, equality can be obtained by setting $\eta_1^{(a)} = \frac{\beta_0^{(a)} - \beta_1^{(a)}}{2}$, which is always allowed under the hypothesis we are working with. The inequality of the third line can always be saturated if we set $\beta_1^{(a)} = \frac{\beta_0^{(a)} - \eta_0^{(a)}}{2}$, also an available choice.

Now, looking at the quantity $\CHSH_{\delta,\epsilon}$ and recalling that $\sum_a p(a) \langle B^0_1B^0_2\rangle_a = 1 - 2\epsilon$, it follows that
\begin{align}
\nonumber
\CHSH_{\delta,\epsilon} & = \sum_a p(a)\CHSH_{\delta}^{(a)}\\
\nonumber
& \leq 4 - 2\,\epsilon - 2 \sum_a p(a) J^{(a)}_{\delta} \\
& \leq 4 - 2\,\epsilon - 2 \sum_a p(a) \left|\beta_0^{(a)} + \eta_0^{(a)} - 1\right|.
\label{Aeq:CHSH_bound}
\end{align}
Let $\tilde{\beta}_0^{(a)} = p(b_1=1|x_1=0,a)$ and $\tilde{\eta}_0^{(a)} = p(b_2=1|x_2=0,a)$, due to normalization, it holds that $\left|\beta_0^{(a)} + \eta_0^{(a)}  - 1\right| = \left| \tilde{\beta}_0^{(a)} + \tilde{\eta}_0^{(a)}  - 1 \right|$. This allows rewriting the last inequality of Eq.\ \eqref{Aeq:CHSH_bound} as
\begin{align}
\CHSH_{\delta,\epsilon} \leq 4 - 2\,\epsilon - \sum_a p(a) \left(\,\left|\beta_0^{(a)} + \eta_0^{(a)}  - 1\right| + \left|\tilde{\beta}_0^{(a)} + \tilde{\eta}_0^{(a)}  - 1 \right|\,\right).
\end{align}

Using the triangle inequality to combine the $a = b_i$ terms and to combine the $a \neq b_i$ terms leads to
\begin{align}
\CHSH_{\delta,\epsilon} \leq 4 - 2\,\epsilon - \left|p(b_1=a|x_1=0) + p(b_2=a|x_2=0) - 1 \right| - \left|p(b_1 = 1-a|x_1=0) + p(b_2 = 1-a|x_2=0) - 1 \right|,
\end{align}
which relates the terms in the moduli to the $\delta$-objectivity inequality \eqref{eq:objcondition} and its complement. The expression in the first modulus satisfies $\sum_i\,\sum_a p(b_i=a, a|x_i) - 1 \geq 1 - 2\delta$, while the expression in the second modulus gives $\sum_i\,\sum_a p(b_i = 1 - a, a|x_i) - 1 \leq 2\delta - 1$. Since $\delta \leq 1/2$, we arrive at
\begin{align}
\CHSH_{\delta,\epsilon} \leq 2 - 2\,\epsilon + 4\,\delta,
\end{align}
which concludes the proof.

\end{proof}

\subsection{Proof of Result 3 and the self-test of the state maximally violating $\mathrm{CHSH}_{0,0}$}
\label{Appendix: self-test}

The idea of performing a self-test of a target state $|\psi\rangle$ from the knowledge about the existence of a given state $|\Psi\rangle$ fulfilling some special property, consists of exploring this property to build a local isometry which, when applied to $|\Psi\rangle$ returns the tensor product of $|\psi\rangle$ with some other ancillary and non-relevant state. Here we will follow similar steps to that shown in section 4 of Ref. \cite{Supic2020}, which is an  up-to-date review of self-testing protocols.

Let $|\Psi^*\rangle$ be a state leading to the maximal quantum violation of the $\CHSH_{0,0}$ inequality when one of the parts measures the observables (and hence Hermitian operators) $B^0_1$ and $B^1_1$ and the other part measures $B^0_2$ and $B^1_2$, i.e.,
\begin{eqnarray}
\label{Aeq: maximal violation}
\langle\Psi^*|B^0_1B^0_2 + B^0_1B^1_2 - B^1_1B^0_2 + B^1_1B^1_2 |\Psi^*\rangle = \frac{5}{2}.
\end{eqnarray}
in which, by hypothesis, it holds that $\langle\Psi^*|B^0_1B^0_2|\Psi^*\rangle = 1$, a condition which entails that
\begin{eqnarray}
\label{Aeq: A0 and B0}
B^0_1|\Psi^*\rangle = B^0_2|\Psi^*\rangle.
\end{eqnarray}

Our first step is to prove the following relation,
\begin{eqnarray}
\label{Aeq: SOS decomposition}
\langle\Psi^*|\left[(B^0_1 - B^0_2)^2 + \frac{1}{2}\left[(B^1_2 - B^0_2) - B^1_1\right]^2\right]|\Psi^*\rangle & = & \langle\Psi^*|\left[\frac{5}{2}\Id - \CHSH_{0,0}\right]|\Psi^*\rangle
\end{eqnarray}

This relation can be achieved via a straightforward calculation using equation \ref{Aeq: A0 and B0} and the condition $\langle\Psi^*|B^0_1 B^0_2|\Psi^*\rangle = 1$:
\begin{eqnarray}
\nonumber
\langle\Psi^*|\left[(B^0_1 - B^0_2)^2 + \frac{1}{2}\left[(B^1_2 - B^0_2) - B^1_1\right]^2\right]|\Psi^*\rangle & = & \langle\Psi^*|\left[(B^0_1)^2 -2B^0_1B^0_2 + (B^0_2)^2 + \frac{1}{2}(B^1_2 - B^0_2)^2 \right.\\
\nonumber
& - & \left. B^1_1(B^1_2 - B^0_2) + \frac{1}{2}(B^1_1)^2\right]|\Psi^*\rangle\\
\nonumber
& = & \langle\Psi^*|\left[2\Id -2B^0_1B^0_2 + \frac{1}{2}((B^1_2)^2 - B^1_2B^0_2 - B^0_2B^1_2 + (B^0_2)^2) \right.\\
\nonumber
& - & \left. B^1_1B^1_2 + B^1_1B^0_2 + \frac{1}{2}\Id\right]|\Psi^*\rangle\\
\nonumber
& = & \langle\Psi^*|\left[\frac{7}{2}\Id -2B^0_1B^0_2 - \frac{1}{2}(B^1_2B^0_2 + B^0_2B^1_2) - B^1_1B^1_2 + B^1_1B^0_2 \right]|\Psi^*\rangle,
\end{eqnarray}
now we use that $\langle\Psi^*|B^0_1 B^0_2|\Psi^*\rangle = 1$ to obtain,
\begin{eqnarray}
\nonumber
\langle\Psi^*|\left[(B^0_1 - B^0_2)^2 + \frac{1}{2}\left[(B^1_2 - B^0_2) - B^1_1\right]^2\right]|\Psi^*\rangle & = & \langle\Psi^*|\left[\frac{5}{2}\Id -B^0_1B^0_2 - \frac{1}{2}(B^1_2B^0_2 + B^0_2B^1_2) - B^1_1B^1_2 + B^1_1B^0_2 \right]|\Psi^*\rangle,
\end{eqnarray}
and then we use Eq. \ref{Aeq: A0 and B0}, from which we get
\begin{eqnarray}
\nonumber
\langle\Psi^*|\left[(B^0_1 - B^0_2)^2 + \frac{1}{2}\left[(B^1_2 - B^0_2) - B^1_1\right]^2\right]|\Psi^*\rangle & = & \langle\Psi^*|\left[\frac{5}{2}\Id -B^0_1B^0_2 - B^0_1B^1_2 + B^1_1B^0_2 - B^1_1B^1_2\right]|\Psi^*\rangle\\
\nonumber
& = & \langle\Psi^*|\left[\frac{5}{2}\Id - \CHSH_{0,0} \right]|\Psi^*\rangle,
\end{eqnarray}
which concludes the proof of Eq. \ref{Aeq: SOS decomposition}.

By combining  \ref{Aeq: SOS decomposition} and eq \ref{Aeq: maximal violation}, we conclude that:
\begin{eqnarray}
\label{Aeq: rel 1}
(B^1_2 - B^0_2)|\Psi^*\rangle = B^1_1|\Psi^*\rangle.
\end{eqnarray}
and using again relation \ref{Aeq: A0 and B0},
\begin{eqnarray}
\label{Aeq: rel 2}
(B^0_1 + B^1_1)|\Psi^*\rangle = B^1_2|\Psi^*\rangle.
\end{eqnarray}

Now, using \ref{Aeq: A0 and B0}, \ref{Aeq: rel 1}, and \ref{Aeq: rel 2},
\begin{align}
\label{Aeq: anticommutators relation}
\nonumber
    \langle\Psi^*|\{B^0_1,B^1_1\}|\Psi^*\rangle & = \langle\Psi^*|B^0_1B^1_1 + B^1_1B^0_1|\Psi^*\rangle\\
    \nonumber
    & = \langle\Psi^*|B^0_1(B^1_2 - B^0_2) + B^1_1B^0_2|\Psi^*\rangle\\
    \nonumber
    & = \langle\Psi^*|(B^1_2 - B^0_2)B^0_1 + B^0_2B^1_1|\Psi^*\rangle\\
    \nonumber
    & = \langle\Psi^*|B^1_2B^0_1 - B^0_2B^0_1 + B^0_2(B^1_2 - B^0_2)|\Psi^*\rangle\\
    \nonumber
    & = \langle\Psi^*|B^1_2B^0_2 - B^0_2B^0_2 + B^0_2B^1_2 - B^0_2B^0_2|\Psi^*\rangle\\
    \nonumber
    & = \langle\Psi^*|B^1_2B^0_2 + B^0_2B^1_2 - 2\Id|\Psi^*\rangle\\
    & = \langle\Psi^*|\{B^0_2,B^1_2\} - 2\Id|\Psi^*\rangle.
\end{align}
in which, because $\langle\Psi^*|[B^0_1,B^1_1]|\Psi^*\rangle = 0$ and $\langle\Psi^*|[B^0_2,B^1_2]|\Psi^*\rangle = 0$, it holds that
\begin{align}
\label{Aeq: bounds on anticommutators}
    \left\{\begin{array}{l}
        \langle\Psi^*|\{B^0_1,B^1_1\}|\Psi^*\rangle = 2\langle\Psi^*|B^1_1B^0_1|\Psi^*\rangle = 2\langle\Psi^*|B^1_1B^0_2|\Psi^*\rangle \\
        \langle\Psi^*|\{B^0_2,B^1_2\}|\Psi^*\rangle = 2\langle\Psi^*|B^0_2B^1_2|\Psi^*\rangle = 2\langle\Psi^*|B^0_1B^1_2|\Psi^*\rangle
    \end{array}\right.,
\end{align}
and so
\begin{align}
\nonumber
    2\langle\Psi^*|B^1_1B^0_2|\Psi^*\rangle & = \langle\Psi^*|2B^0_1B^1_2 - 2\Id|\Psi^*\rangle\\
    \nonumber
    & \implies \langle\Psi^*|B^0_1B^1_2 - B^1_1B^0_2|\Psi^*\rangle = 1.
\end{align}
Because of the constraint $\langle\Psi^*|B^0_1B^0_2|\Psi^*\rangle = 1$ and the Eq. \ref{Aeq: maximal violation}, this means that
\begin{align}
\nonumber
    \frac{1}{2} & = \langle\Psi^*|B^1_1B^1_2|\Psi^*\rangle\\
    \nonumber
    & = \langle\Psi^*|B^1_1(B^0_1 + B^1_1)|\Psi^*\rangle\\
    \nonumber
    & = \langle\Psi^*|B^1_1B^0_1 + \Id|\Psi^*\rangle\\
    \nonumber
    & \implies \langle\Psi^*|B^1_1B^0_1 |\Psi^*\rangle = -\frac{1}{2}.
\end{align}
Going back to Eq. \ref{Aeq: bounds on anticommutators},
\begin{align}
\label{Aeq: antic. 1}
    \langle\Psi^*|\{B^0_1,B^1_1\}|\Psi^*\rangle = -1\;\;\; \implies\;\;\; \{B^0_1,B^1_1\}|\Psi^*\rangle = -|\Psi^*\rangle,
\end{align}
which can be replaced in Eq. \ref{Aeq: anticommutators relation} to obtain
\begin{align}
\label{Aeq: antic. 2}
    \langle\Psi^*|\{B^0_2,B^1_2\}|\Psi^*\rangle = 1\;\;\; \implies\;\;\; \{B^0_2,B^1_2\}|\Psi^*\rangle = |\Psi^*\rangle
\end{align}

Now let us define the following operators:
\begin{align}
\nonumber
    Z_1\equiv B^0_1 & \;\;\;\;\;\;\;\;\;\;\; & X_1 \equiv -\frac{\sqrt{3}}{3}\left(2B^1_1 + B^0_1\right)\\
    Z_2\equiv B^0_2 & & X_2 \equiv \frac{\sqrt{3}}{3}\left(B^0_2 - 2B^1_2\right).
\end{align}
These operators are such that,
\begin{eqnarray}
Z_1|\Psi^*\rangle = Z_2|\Psi^*\rangle,
\end{eqnarray}
from Eq. \ref{Aeq: A0 and B0}. Additionally, from eqs.\eqref{Aeq: rel 1}, and \eqref{Aeq: rel 2}:
\begin{eqnarray}
\label{Aeq: X_1 =ish X_2}
\nonumber
X_1|\Psi^*\rangle & = &  -\frac{\sqrt{3}}{3}\left(2B^1_1 + B^0_1\right)|\Psi^*\rangle\\
\nonumber
& = & -\frac{\sqrt{3}}{3}\left(2(B^1_2 - B^0_2) + B^0_2\right)|\Psi^*\rangle\\
\nonumber
& = & -\frac{\sqrt{3}}{3}\left(2B^1_2 - B^0_2\right)|\Psi^*\rangle\\
& = & X_2|\Psi^*\rangle,
\end{eqnarray}

Also, the effect of the anticommutator of $Z_1$ and $X_1$, $\{Z_1,X_1\}$, and the anticommutator of $Z_2$ and $X_2$ are related when acting on $|\Psi^*\rangle$:
\begin{align}
\nonumber
    \{Z_1,X_1\}|\Psi^*\rangle & = -\frac{\sqrt{3}}{3}\left(2B^0_1B^1_1 + B^0_1B^0_1 + 2B^1_1B^0_1 + B^0_1B^0_1\right)|\Psi^*\rangle\\
    \nonumber
    & = -\frac{\sqrt{3}}{3}\left(2\{B^0_1,B^1_1\}+ 2\Id\right)|\Psi^*\rangle\\
    & = 0
\end{align}
and similarly,
\begin{align}
\nonumber
    \{Z_2,X_2\}|\Psi^*\rangle & = \frac{\sqrt{3}}{3}\left(B^0_2B^0_2 - 2B^0_2B^1_2 + B^0_2B^0_2 - 2B^1_2B^0_2\right)|\Psi^*\rangle\\
    \nonumber
    & = \frac{\sqrt{3}}{3}\left(-2\{B^0_2,B^1_2\}+ 2\Id\right)|\Psi^*\rangle\\
    & = 0
\end{align}
in which we used relations \ref{Aeq: antic. 1} and \ref{Aeq: antic. 2}.

Furthermore, $Z_1$ and $Z_2$ are unitary by construction, and $X_1$ and $X_2$ act in $|\Psi^*\rangle$ as a unitary operator:
\begin{align}
\nonumber
    X_1^{\dagger}X_1|\Psi^*\rangle & = \frac{1}{3}\left(2B^1_1 + B^0_1\right)\left(2B^1_1 + B^0_1\right)|\Psi^*\rangle\\
    \nonumber
    & = \frac{1}{3}\left(4B^1_1B^1_1 + 2B^1_1B^0_1 + 2B^0_1B^1_1 + B^0_1B^0_1\right)|\Psi^*\rangle\\
    \nonumber
    & = \frac{1}{3}\left(5\Id + 2(B^1_1B^0_1 + B^0_1B^1_1)\right)|\Psi^*\rangle\\
    \nonumber
    & = \frac{1}{3}\left(5\Id - 2\Id\right)|\Psi^*\rangle\\
    & = |\Psi^*\rangle.
\end{align}
The proof for $X_2$ follows from equation \ref{Aeq: X_1 =ish X_2}. Finally, as the sub-indices on the B's label different parties, it also holds true that $[Z_1,Z_2]=0$ as well as $[X_1,X_2]=0$.

Now consider the local isotropic transformation depicted in Fig. \ref{fig: Circuit}. The final state is dubbed $|\Psi_{final}\rangle$ and given by:
\begin{figure}
    \centering
    \includegraphics[scale =0.3]{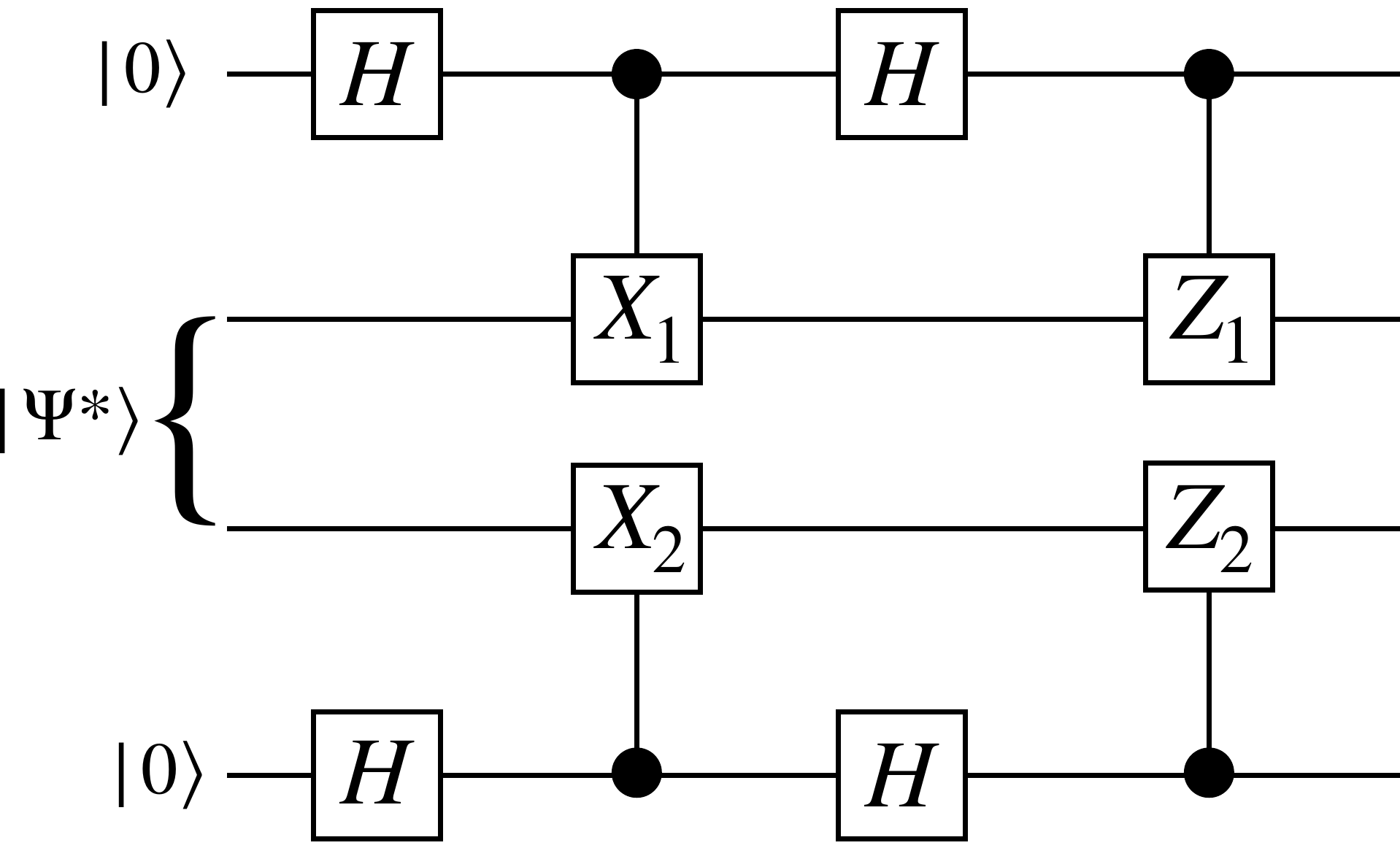}
    \caption{The partial swap circuit, a local isotropic transformation usually employed in self-testing protocols.}
    \label{fig: Circuit}
\end{figure}
\begin{align}
    |\Psi_{final}\rangle = \frac{1}{4}\sum_{j,k = 0}^1|jk\rangle\otimes\left[Z_1^j(\Id + (-1)^jX_1)Z_2^k(\Id + (-1)^kX_2)|\Psi^*\rangle\right].
\end{align}
Because of relation \ref{Aeq: X_1 =ish X_2} and also because $X_1$ commutes with $X_2$, all terms for which $j+k=1$ vanish, leading to
\begin{align}
    |\Psi_{final}\rangle = \frac{1}{4}|00\rangle\otimes\left[(\Id + X_1)(\Id + X_2)|\Psi^*\rangle\right] + \frac{1}{4}|11\rangle\otimes\left[Z_1(\Id - X_1)Z_2(\Id - X_2)|\Psi^*\rangle\right],
\end{align}
in which, we can use the fact that $Z_1$ and $X_1$, as well as $Z_2$ and $X_2$, anticommute to perform the following calculation,
\begin{align}
\nonumber
    Z_1(\Id - X_1)Z_2(\Id - X_2)|\Psi^*\rangle & = (Z_1 - Z_1X_1)(Z_2 - Z_2X_2)|\Psi^*\rangle\\
    \nonumber
    & = (Z_1 + X_1Z_1)(Z_2 + X_2Z_2)|\Psi^*\rangle\\
    \nonumber
    & = (\Id + X_1)Z_1(\Id + X_2)Z_2|\Psi^*\rangle\\
    \nonumber
    & = (\Id + X_1)(\Id + X_2)Z_1Z_2|\Psi^*\rangle,\\
    & = (\Id + X_1)(\Id + X_2)Z_1Z_1|\Psi^*\rangle,\\
    & = (\Id + X_1)(\Id + X_2)\Id|\Psi^*\rangle,
\end{align}
which leads to 
\begin{align}
\nonumber
    |\Psi_{final}\rangle &= \frac{1}{4}|00\rangle\otimes\left[(\Id + X_1)(\Id + X_2)|\Psi^*\rangle\right] + \frac{1}{4}|11\rangle\otimes\left[(\Id + X_1)(\Id + X_2)|\Psi^*\rangle\right]\\
    \nonumber
    & = \frac{1}{\sqrt{2}}\left(|00\rangle + |11\rangle\right)\otimes\left[\frac{1}{2\sqrt{2}}(\Id + X_1)(\Id + X_2)|\Psi^*\rangle\right]\\
    \nonumber
    & = |\phi^+\rangle\otimes|\zeta_{junk}\rangle,
\end{align}
thus completing our self-test.

\subsection{A semi-definite program formulation of the problem}
\label{app:sdp}

Given the distribution $p(b_1,b_2|x_1,x_2)$ one may search for the least $\delta$ necessary to describe it with a $\delta$-objective distribution $p(a,b_1,b_2|x_1,x_2)$. Using a general distribution, this is an optimization that can be realized with a linear program, the restriction to quantum attainable distributions $p(a,b_1,b_2|x_1,x_2)$, however, is a hard task in general. Ideally, the objective would be finding the best combination of density operator $\rho_{A B_1 B_2}$ and POVMs $M_A^{(a)}$, $M^{(b_1)}_{B_1,\,x_1}$ and $M^{(b_2)}_{B_2,\,x_2}$ such that
\begin{equation}
    p(a,b_1,b_2|x_1,x_2) = \tr\left( \rho_{A B_1 B_2} M_A^{(a)} \otimes M^{(b_1)}_{B_1,\,x_1} \otimes M^{(b_2)}_{B_2,\,x_2} \right),\,\forall a,b_1,b_2,x_1,x_2,
\end{equation}
satisfying constraint \eqref{eq:objcondition} with the lowest value possible for $\delta$. Since no restriction is made on the dimensions of the involved Hilbert spaces, these would also become parameters to be included in the optimization. Upper bounds on the optimal $\delta$ can be done simply by providing a generic realization for $p(a,b_1,b_2|x_1,x_2)$ that is not necessarily the optimal one (noticing that we are minimizing $\delta$). A way to lower bound the optimal $\delta$ can be realized by relaxing the problem and imposing some but not all necessary conditions that any quantum-attainable distribution should satisfy.

A standard method in the literature is to employ the NPA hierarchy \cite{NPA_2007} of semidefinite program tests, which is proven to converge to the set of quantum distributions \cite{NPA_2008}. The method proposes to further constrain the possible distributions $p(a,b_1,b_2|x_1,x_2)$ by requiring that it should also be compatible with a matrix of statistical moments that can always be made positive semidefinite whenever there exists a quantum realization for $p$. For a given $p$ with unknown realization, the moment matrix is built up by considering free variables and the constraints implied if a quantum realization exists: a sequence of observables $O_1,\ldots,O_n$ and a quantum state $|\psi\rangle_{A B_1 B_2}$ are assumed, which can be chosen pure by using a high enough dimension for the Hilbert space, and each entry in the moment matrix $\Gamma$ is given by $\Gamma_{ij} = \langle\psi| O_i^\dagger O_j |\psi\rangle$. In general, the observables are chosen to be the measurement operators $M_A^{(a)}$, $M^{(b_1)}_{B_1,\,x_1}$ and $M^{(b_2)}_{B_2,\,x_2}$ (which can also be assumed projective by using a dimension big enough) and strings formed by them, with length up to a given $k \geq 1$. In some cases, the entries can be associated with components of the distribution $p(a,b_1,b_2|x_1,x_2)$, or they can be identified with other entries. If a quantum realization for $p$ exists, $k$ can be taken to an arbitrarily high value; failure to obtain a positive semidefinite moment matrix $\Gamma$ for a given $k$ then certificates that the behavior is post-quantum.

Employing this method to bound a distribution $p(a,b_1,b_2|x_1,x_2)$ which reduces to an observable distribution $p(b_1,b_2|x_1,x_2)$ with specific values for the CHSH expression, allows us to bound the objectivity that can be associated to the observed values when $x_1=0$ or $x_2=0$. In Fig.\ \ref{fig:Delta_CHSH} we show different curves for the minimum $\delta$ as a function of the CHSH value, considering also different agreements between the observers, given by the extra (linear) constraint $\sum_{b_1} p(b_1=b_2|x_1=0,\,x_2=0) = 1 - \epsilon$, with $0 \leq \epsilon \leq 0.5$.

\subsection{Proof of result 4}
\label{app:result4}

To prove Result \ref{result 4}, we are going to show that it is always possible to define a classical joint distribution $p(a_1,a_2)$ in a way that $(a_1,a_2)$ plays the rule of a hidden-variable in a classical model that can explain any distribution $p(b_1,b_2|x_1,x_2)$ obtained in an experiment with quantum systems.

First of all, we take advantage of the fact that we are not making any constrain in the dimension of the systems under consideration, to work only with pure states. So consider a pure state $|\Psi\rangle \in (\mathcal{H}_1\otimes\mathcal{H}_2)$. Let $\{M^{(1)}_{0|0},M^{(1)}_{1|0},M^{(1)}_{0|1},M^{(1)}_{1|1},M^{(2)}_{0|0},M^{(2)}_{1|0},M^{(2)}_{0|1},M^{(2)}_{1|1}\}$ be a set of projectors, such that $\sum_{b=0}^1M^{(i)}_{b,x} = \Id$ for $i,x\in\{0,1\}$, $M^{(i)}_{b,x}\in\mathcal{L}(\mathcal{H}_i)$, and
\begin{eqnarray}
M^{(i)}_{b,x} = \sum_{k}|\phi^{(i)}_{k,b,x}\rangle\langle \phi^{(i)}_{k,b,x}|
\end{eqnarray}
in which
\begin{eqnarray}
\langle\phi^{(i)}_{k,b,x}| \phi^{(i)}_{k',b',x}\rangle = \delta_{b,b'}\delta_{k,k'}.
\end{eqnarray}

The condition $p(b_1=b_2|x_1=x_2)$ implies that $E_{00} = E_{11} = 1$ and hence, no symmetry of the CHSH having these two terms with different signs can be violated. It also implies that,
\begin{eqnarray}
|\Psi\rangle & = & \sum_{k, l}\alpha_{k,l} |\phi^{(0)}_{k,0,0}\rangle\otimes|\phi^{(1)}_{l,0,0}\rangle + \sum_{r, s}\beta_{r,s} |\phi^{(0)}_{r,1,0}\rangle\otimes|\phi^{(1)}_{s,1,0}\rangle
\end{eqnarray}
and, at the same time,
\begin{eqnarray}
|\Psi\rangle & = & \sum_{p, q}\alpha'_{p,q} |\phi^{(0)}_{p,0,1}\rangle\otimes|\phi^{(1)}_{q,0,1}\rangle + \sum_{u, v}\beta'_{u,v} |\phi^{(0)}_{u,1,1}\rangle\otimes|\phi^{(1)}_{v,1,1}\rangle.
\end{eqnarray}

Now, define $\tilde{p}(a_0,a_1)$ as follows,
\begin{eqnarray}
\nonumber
\tilde{p}(a_0,a_1) & \equiv & p(b_1=a_0,b_2=a_1|x_1 = 0,x_2 = 1)\\
\nonumber
& = & \langle\Psi|M_{a_0|0}^{(0)}\otimes M_{a_1|1}^{(1)}|\Psi\rangle\\
\nonumber
& = & \langle\Psi|\left[\left(\sum_{j}|\phi^{(0)}_{j,a_0,0}\rangle\langle \phi^{(0)}_{j,a_0,0}|\right)\otimes \left(\sum_{k}|\phi^{(1)}_{k,a_1,1}\rangle\langle \phi^{(1)}_{k,a_1,1}|\right)\right]\left( \sum_{p, q}\alpha'_{p,q} |\phi^{(0)}_{p,0,1}\rangle\otimes|\phi^{(1)}_{q,0,1}\rangle + \sum_{u, v}\beta'_{u,v} |\phi^{(0)}_{u,1,1}\rangle\otimes|\phi^{(1)}_{v,1,1}\rangle\right)\\
\nonumber
& = & \langle\Psi|\left[\left(\sum_{j}|\phi^{(0)}_{j,a_0,0}\rangle\langle \phi^{(0)}_{j,a_0,0}|\right)\otimes\Id^{(1)}\right] \left(\sum_{k,p,q}\left(\alpha'_{pq}\delta_{a_1,0} + \beta'_{p,q}\delta_{a_1,1}\right)\delta_{k,q}|\phi_{p,a_1,1}^{(0)}\rangle\otimes|\phi^{(1)}_{k,a_1,1}\rangle\right)\\
\nonumber
& = & \left(\sum_{u, v}\alpha_{u,v}^* \langle\phi^{(0)}_{u,0,0}|\otimes\langle\phi^{(1)}_{v,0,0}| + \sum_{r, s}\beta_{r,s}^* \langle\phi^{(0)}_{r,1,0}|\otimes\langle\phi^{(1)}_{s,1,0}|\right)\left[\left(\sum_{j}|\phi^{(0)}_{j,a_0,0}\rangle\langle \phi^{(0)}_{j,a_0,0}|\right)\otimes\Id^{(1)}\right] \\
\nonumber
& \times & \left(\sum_{k,p,q}\left(\alpha'_{pq}\delta_{a_1,0} + \beta'_{p,q}\delta_{a_1,1}\right)\delta_{k,q}|\phi_{p,a_1,1}^{(0)}\rangle\otimes|\phi^{(1)}_{k,a_1,1}\rangle\right)\\
\nonumber
& = & \left(\sum_{j,u,v}\left(\alpha_{u,v}^*\delta_{a_0,0} + \beta_{u,v}^*\delta_{a_0},1\right)\delta_{v,j}\langle \phi^{(0)}_{j,a_0,0}|\otimes\langle\phi^{(1)}_{j,a_0,0}|\right)\left(\sum_{k,p}\left(\alpha'_{pq}\delta_{a_1,0} + \beta'_{p,q}\delta_{a_1,1}\right)|\phi_{p,a_1,1}^{(0)}\rangle\otimes|\phi^{(1)}_{k,a_1,1}\rangle\right)\\
\nonumber
& = & \sum_{j,u,k,p}\left(\alpha_{u,v}^*\delta_{a_0,0} + \beta_{u,v}^*\delta_{a_0},1\right)\left(\alpha'_{pq}\delta_{a_1,0} + \beta'_{p,q}\delta_{a_1,1}\right)\langle \phi^{(0)}_{j,a_0,0}|\phi_{p,a_1,1}^{(0)}\rangle\langle\phi^{(1)}_{j,a_0,0}|\phi^{(1)}_{k,a_1,1}\rangle\\
\nonumber
& = & \left(\langle\Psi|M_{a_1|1}^{(0)}\otimes M_{a_0|0}^{(1)}|\Psi\rangle\right)^*\\
\nonumber
& = & \langle\Psi|M_{a_1|1}^{(0)}\otimes M_{a_0|0}^{(1)}|\Psi\rangle\\
\nonumber
& = & p(b_1=a_1,b_2=a_0|x_1=1,x_2=0)
\end{eqnarray}

Hence, given a quantum distribution (which always satisfies the NS condition) obeying $p(b_1=b_2|x_1=x_2) = 1$ one can always define a distribution $\tilde{p}(a_0,a_1)$ such that, by construction $\tilde{p}(a_0,a_1|x_1,x_2) = \tilde{p}(a_0,a_1)$, i.e., it satisfies the NSD condition. More precisely, for any quantum distribution obeying $p(b_1=b_2|x_1=x_2) = 1$, it holds that,
\begin{eqnarray}
\label{Aeq: p(b_1,b_2|x_1,x_2)}
p(b_1,b_2|x_1,x_2) = \left\{\begin{array}{ll}
    \delta_{b_1,b_2}\tilde{p}(a_0=b_1) &\mbox{, if }x_1=0,\mbox{ and } x_2=0  \\
    \tilde{p}(a_0=b_1,a_1=b_2) &\mbox{, if }x_1=0,\mbox{ and } x_2=1  \\
    \tilde{p}(a_1=b_1,a_0=b_2) &\mbox{, if }x_1=1,\mbox{ and } x_2=0 \\
    \delta_{b_1,b_2}\tilde{p}(a_1=b_1) &\mbox{, if }x_1=1,\mbox{ and } x_2=1
\end{array}\right.,
\end{eqnarray}
 in which we have just proven the second and the third equations, and one can obtain the first and the fourth by applying the NS condition combined with the imposition $p(b_1=b_2|x_1=x_2) = 1$ to the second and third equations.
 
 Another way of writing \ref{Aeq: p(b_1,b_2|x_1,x_2)} is as follows
 \begin{align}
 \label{Aeq: LHV model}
     p(b_1,b_2|x_1,x_2) = \sum_{a_1,a_2}\delta_{b_1,f_{a_1,a_2}(x_1)}\delta_{b_2,f_{a_1,a_2}(x_2)}p(a_1,a_2),
 \end{align}
 in which
 \begin{align}
     f_{a_1,a_2}(x) = a_x.
 \end{align}
 By definition, eq \ref{Aeq: LHV model} is a local hidden-variable model for $p(b_1,b_2|x_1,x_2)$, implying that this distribution cannot violate the CHSH inequality, as announced in Result \ref{result 4}.

\subsection{Mapping the experimental scheme to the Darwin scenario}
\label{sec:mapping}
In section~\ref{sec:setup} we described the dynamics of the system treating independently the temporal modes of the two photons to simplify the presentation.
Here we will describe more in detail the evolution of the system and environment in terms of their joint temporal mode.
In the experimental scheme, two photons, generated by a 
collinear SPDC (Spontaneous Parametric Down-Conversion) process of type II, in a ppKTP (Periodically-Poled Potassium Titanyl Phosphate) crystal (see section~\ref{sec:setup}), interact with a birefringent crystal in one of the branches of the Sagnac interferometer. 
In this interaction, the polarization and the temporal mode of the two photons, get entangled. 
The strength of the interaction can be tuned by enlarging the thickness of the birefringent plate.
In order to map the scheme into the quantum Darwinism scenario described in section~\ref{sec:darwinism}, we identify the polarization of the two photons after they leave the interferometer as
two fragments of the environment, identified by the four-dimensional Hilbert space $\mathcal{H}_{B_1}\otimes \mathcal{H}_{B_2}$.
On the other hand, the joint temporal mode of the two photons represents the quantum system probed by the environment, corresponding to the Hilbert space $H_A$.
A joint temporal state of the two photons generated in our scheme can be written, in general, in the following form:
\begin{equation}
    \ket{f_{HV}} = \int d \omega_H d \omega_V \tilde f(\omega_H, \omega_V) a^\dagger_{H, \omega_H}a^\dagger_{V, \omega_V} \ket{0} =
    \int d t_H d t_V f(t_H, t_V) A^\dagger_{H, t_H} A^\dagger_{V, t_V} \ket{0} \; ,
\end{equation}
where $a^\dagger_{H, \omega_H}, a^\dagger_{V, \omega_V}$ are the creation operators for photons with horizontal and vertical polarization respectively, while $A^\dagger_{P, t}$ is defined as
\begin{equation}
    A^\dagger_{P, t} = \int \frac{d \omega}{2\pi} e^{-i\omega t} a^\dagger_{P, \omega} \; .
\end{equation}
If we consider a monochromatic pump with $\omega_p$, we have that the joint spectral amplitude $\tilde f(\omega_H, \omega_V) = \delta(\omega_p - \omega_H - \omega_V) \tilde f'(\omega_H, \omega_V)$, ($ \tilde f'(\omega_H, \omega_V)$ corresponds to the phase matching function in the SPDC process \cite{boyd2020nonlinear})  and its Fourier transform, i.e. the joint temporal function, will depend only on the difference $\Delta t = t_H - t_V$, a part from a global phase shift $f(t_H, t_V) = e^{i\omega_p (t_H + t_V)/2} g(t_H - t_V)$.

The birefringent crystal will change this temporal mode by introducing a shift $t_H \rightarrow t_H + \tau$.
The overlap with the original time mode is then given by
\begin{equation}
    \Delta = \braket{f_{HV}}{f_{HV}^\tau} = 
    e^{i \omega_p \tau /2}\int d \Delta t \; g^*(\Delta t) g(\Delta t + \tau), 
\end{equation}
where $\ket{f_{HV}^\tau}$ is the state with the translated temporal mode $f(t_H+ \tau, t_V )$.
Using the notation just introduced we can describe the state in the interferometer, after the birefringent crystal, as
\begin{equation}
    \frac{1}{\sqrt{2}}\left(e^{-i\phi}\ket{f_{HV}, \mathrm{CW}} + \ket{f_{HV}^\tau, \mathrm{CCW}}\right) = 
    \frac{1}{\sqrt{2}}\left(e^{-i\phi}\ket{f_{HV}, \mathrm{CW}} + \Delta \ket{f_{HV}, \mathrm{CCW}} + \gamma \ket{f_{HV}^\perp, \mathrm{CCW}}\right)
\end{equation}
where $\mathrm{CCW}, \mathrm{CW}$ represent the counter-clockwise and clockwise orientations inside the interferometer, and $\gamma \ket{f_{HV}^\perp} = (\mathds{I} - \ket{f_{HV}}\bra{f_{HV}}) \ket{f_{HV}^\tau}$ is the projection of $\ket{f_{HV}^\tau}$ on the space orthogonal to $\ket{f_{HV}}$, with $\gamma = \sqrt{1 - |\Delta|^2}$.
The phase $\phi$ can be appropriately tuned using a liquid crystal (not shown in Fig~\ref{fig:piatte}).
If we now associate a qubit $\ket{0}_A, \ket{1}_A$ to $\ket{f_{HV}}$ and its orthogonal state, respectively, we end up, after the interferometer, with the state
\begin{equation}
    \frac{1}{\sqrt{2}} \left( 
    e^{-i\phi}\ket{V}_{B_1}\ket{H}_{B_2} + \Delta \ket{H}_{B_1}\ket{V}_{B_2} 
    \right) \ket{0}_A + \frac{\gamma}{\sqrt{2}} \ket{H}_{B_1}\ket{V}_{B_2} \ket{1}_A \; ,
\end{equation}
which, after tracing out $A$, corresponds exactly to the state \eqref{eq:finalstate}.

This shows that the internal degree of freedom probed by the environment is precisely the \emph{joint} temporal mode of the two photons and, in particular, the mode $\ket{f_{HV}}$ and its orthogonal complement.

\subsection{Ab-initio optimization protocol}
\label{sec:abinitio}

The gradient-free optimization algorithm employed in our proof-of-principle experiment is based on the \textit{Stochastic Nelder-Mead (SNM)} introduced in Ref. \cite{chang2012stochastic}, considered an improvement over the well-known \textit{Nelder-Mead} algorithm when dealing with noisy cost functions $F(\vec{x}) = f(\vec{x}) + \mathcal{N}(\vec{x})$. In particular, it tends to avoid getting stuck in a local optimum of the function provided to the algorithm, which could be due to statistical fluctuations in the noise term $\mathcal{N}(\vec{x})$. This algorithm has been recently used for the ab-initio optimization of the violation of Bell inequalities and randomness generation, extremizing the device-independent approach in experimental tasks \cite{poderini2022ab}. In this approach, the experimental apparatus is treated as a black box, that is, the action of the controllable parameters (see Fig. \ref{fig:abinitio} a) is unknown to the algorithm and only the noisy output statistics is used to reach the optimal value of a given cost function, that in our case is composed of the agreement between the observers and subsequently the CHSH parameter.

\begin{figure}
    \centering
    \includegraphics[scale=0.80]{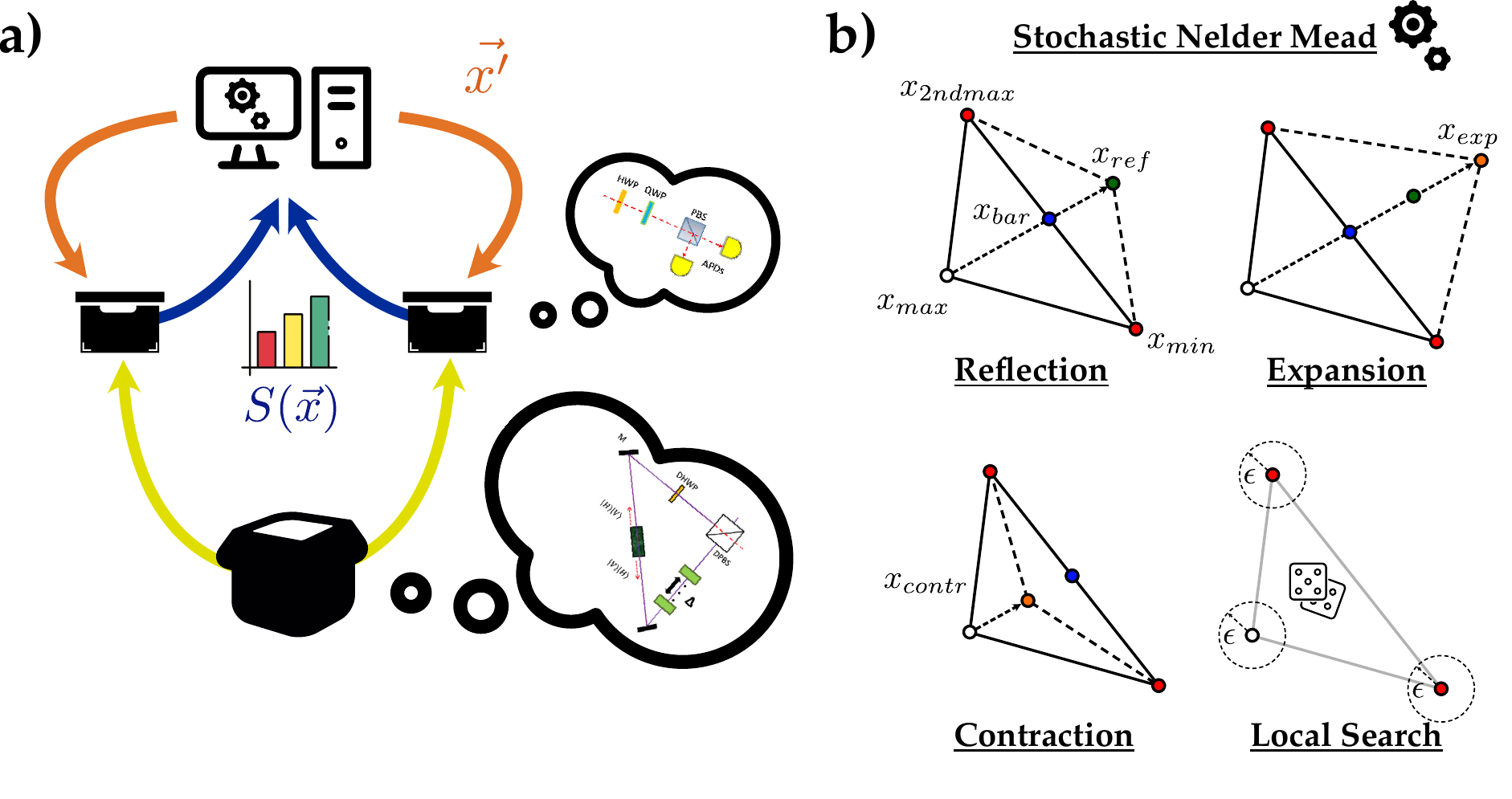}
    \caption{$\bm{a)}$ Conceptual scheme of the ab-initio optimization process in an experimental scenario. Both the photonic system being produced and the measurement stations performing polarization measurement are unknown to the algorithm, i.e. they are treated as black boxes. Nonetheless, the goal of the algorithm will be to find the maximal value of a noisy function $S(\vec{x})$ - e.g. the CHSH quantity - tuning progressively the parameters $\vec{x}$, which have an apriori unknown relation with the measurements which are being carried out. In this case, the parameters $\vec{x}$ have a one-to-one correspondence with the rotation of two motorized waveplates. 
    $\bm{b)}$ Pictorial representation of possible evolutions of the geometrical simplex tracked by the SNM algorithm as described  carried out by the SNM algorithm in Appendix G. First, the barycenter $x_{bar}$ of the simplex is computed, together with a reflection point $x_{ref}$. If such a point is promising, i.e. $F(x_{ref}) \leq F(x_{min})$, an expansion point $x_{exp}$ is computed. On the contrary, if $F(x_{max}) \leq F(x_{ref})$, a contraction point $x_{contr}$ is tested. If such a step fails to find a point such that $F(x_{contr}) \leq F(x_{max})$, an Adaptive Random Search is performed. With probability $1-P$, a local search occurs: a new point $x_{RNG}$ is uniformly sampled in a hypersphere of radius $\epsilon$ centered in a randomly chosen simplex point.}
    \label{fig:abinitio}
\end{figure}

The SNM algorithm performs the optimization of a given cost function over a d-dimensional parameter space, finding
\begin{equation}
\min_{\vec{x}} F(\vec{x}) \quad \mathrm{where} \quad F(\vec{x}): \mathbb{R}^d \rightarrow \mathbb{R}  ,
\end{equation}
starting from an initial (d+1)-dimensional \textit{simplex} of points $\Sigma_0 = \{x_0, \dots x_d\}$, sampled in a given parameter range through the \textit{Latin Hypercube Sampling} algorithm \cite{chang2012stochastic}. 

The algorithm flow proceeds as follows (see also  Fig. \ref{fig:abinitio} b for a graphical representation of the simplex evolutions):

\begin{enumerate}
    \setcounter{enumi}{-1}
    \item At first, the cost function $F(\cdot)$ is evaluated in each point of the initial simplex $\Sigma_0$. 
    \item If $\dim(\Sigma_k) > d + 1$, the point in which the cost function assumes the highest value is discarded. 
    \item The remaining points of the simplex $\Sigma_k$ are ranked with respect to the value of the cost function $F(x_i)$. $x_{max}$, $x_{2ndmax}$, $x_{min}$; i.e. the points in which the cost function assumes the highest, the second highest and the lowest values, are identified.

    \item The barycenter of the simplex points is computed as $$ x_{bar} = \sum_{x \in \Sigma_k \setminus x_{max}} x/d, $$
    and the \textit{reflection} point is computed as
    $$x_{ref} = (1 + \alpha) x_{bar} - \alpha x_{max}; \quad \alpha > 0.$$
    The cost function $F(x_{ref})$ is evaluated. If $F(x_{min}) \leq F(x_{ref}) < F(x_{2ndmax})$, then $\Sigma_{k+1} = \Sigma_k \cup x_{ref}$. Else:

    \begin{enumerate}
        \item If $F(x_{ref}) \leq F(x_{min})$, further expand the simplex in the direction of $x_{ref}$, computing an \textit{expansion} point:
        $$x_{exp} = (1 - \gamma) x_{bar} + \gamma x_{ref}; \quad \gamma > 1.$$
        Compute the cost function in $x_{exp}$: if $F(x_{exp}) < F(x_{ref})$ then $\Sigma_{k+1} = \Sigma_k \cup x_{exp}$, else $\Sigma_{k+1} = \Sigma_k \cup x_{ref}$. Go back to step 1.
        \item If $F(x_{2ndmax}) \leq F(x_{ref}) < F(x_{max})$, perform an \textit{external} contraction:
        $$x_{contr} = (1 - \beta) x_{bar} + \beta x_{ref}; \quad 0 \leq \beta \leq 1.$$
        Compute the cost function in the contracted point $x_{contr}$: if $F(x_{contr}) \leq F(x_{ref})$, impose $\Sigma_{k+1} = \Sigma_k \cup x_{contr}$ and go back to step 1.
        \item If instead $F(x_{max}) \leq F(x_{ref})$, perform an \textit{internal} contraction:
        $$x_{contr} = (1 - \beta) x_{bar} + \beta x_{max}; \quad 0 \leq \beta \leq 1.$$
        Compute the cost function in the contracted point $x_{contr}$: if $F(x_{contr}) \leq F(x_{max})$, impose $\Sigma_{k+1} = \Sigma_k \cup x_{contr}$ and go back to step 1.
    \end{enumerate}

    \item If the \textit{contraction} fails, perform an \textit{Adaptive Random Search (ARS)}, stochastically sampling the parameter space:
    \begin{enumerate}
        \item With probability $P$, a \textit{global} search is performed, choosing a point $x_{RNG}$ uniformly sampling the parameter space within given bounds for each parameter.
        \item With probability $1-P$, a \textit{local} search is performed: a point of the simplex $x_i$ is randomly chosen and $x_{RNG}$ is uniformly sampled within an hypersphere of radius $\epsilon$ centered in $x_i$.
    \end{enumerate}
    The adaptive random search is performed until a point such that $F(x_{RNG}) < F(x_{max})$ is found. Impose $\Sigma_{k+1} = \Sigma_k \cup x_{RNG}$ and go back to step 1.
\end{enumerate}

Within our experimental approach, performing polarization measurements of single photon pairs, each observable $B^i_j$ is associated to two parameters $(\delta^i_j, \phi^i_j)$ corresponding to the rotation angles of the optical axes of the two waveplates implementing, together the PBS and single-photon detectors, such measurements. Albeit the SNM algorithms has no knowledge of this physical equivalence, the parameters on which it has to optimize a given cost function will have a one-to-one correspondence to a set of angles $(\delta^i_j, \phi^i_j)$ which are then provided to the waveplates' motorized rotation mounts, while expectation values $\mean{B^k_1B^l_2}$ are obtained by recording photon-coincidence counts on single photon avalanche photodiodes within a window of $2.4$ ns.
Moreover, note that in our proof-of-principle experiment, we need to perform a \textit{constrained} optimization of the CHSH parameter:
\begin{equation}
    \mathrm{CHSH} =\mean{B^0_1B^0_2} + \mean{B^0_1B^1_2} \\- \mean{B^1_1B^0_2} + \mean{B^1_1B^1_2}
\end{equation}
under the constraint $\mean{B^0_1B^0_2}=1-2\epsilon$, for a given $\epsilon$. 
To address this problem, we perform first an optimization of the agreement between the two observers and then the optimization of the CHSH value, according to the following two-step process:
\begin{enumerate}
    \item First, fix a value of $\epsilon$ and perform the 4-parameter minimization of the function:
    $$A = \abs{\mean{B^0_1B^0_2} - (1-2\epsilon)}$$
    The minimization is performed for 100 proposed points, with early stopping if $A < 0.03$.
    \item Keeping fixed the parameters related to $\{B^0_1, B^0_2\}$, perform a minimization of:
    $$\mathrm{S} = - \abs{\mean{B^0_1B^0_2} + \mean{B^0_1B^1_2} \\- \mean{B^1_1B^0_2} + \mean{B^1_1B^1_2}}$$
    Extract $S$ and $\epsilon = \frac{1 - \mean{B^0_1B^0_2}}{2}$ from the best point in the simplex after 250 explored points in the 4-dimensional parameter space corresponding to $\{B^1_1, B^1_2\}$. 
\end{enumerate}

\end{widetext}
\end{document}